\documentclass[12pt]{article}
\usepackage{a4wide}
\usepackage{latexsym}
\usepackage{amsfonts}
\usepackage{amssymb}

\def\l{\langle}
\def\r{\rangle} 
\def\bq{\begin{eqnarray}}
\def\eq{\end{eqnarray}}

\def\bs{\begin{small}}
\def\es{\end{small}}
\def\eps{\varepsilon}

\begin{document}

\thispagestyle{empty}

\begin{flushright}
UPRF-03-014
\end{flushright}

\vspace{1.5cm}

\begin{center}
  {\Large \bf Subtraction terms for one-loop amplitudes with one unresolved parton}\\[.3cm]
  \vspace{1.7cm}
  {\sc Stefan Weinzierl$^{1}$}\\
  \vspace{1cm}
  {\it Dipartimento di Fisica, Universit\`a di Parma,\\
       INFN Gruppo Collegato di Parma, 43100 Parma, Italy} \\
\end{center}

\vspace{2cm}

\begin{abstract}\noindent
  {
   Fully differential next-to-next-to-leading order calculations require
   a method to cancel infrared singularities.
   In a previous publication, I discussed the general setup for the 
   subtraction method at NNLO.
   In this paper I give all subtraction terms for electron-positron annihilation 
   associated with one-loop amplitudes with one unresolved parton.
   These subtraction terms are integrated within dimensional regularization over the
   unresolved one-particle phase space.
   The results can be used with all variants of dimensional regularization
   (conventional dimensional regularization, the 't Hooft-Veltman scheme and the
   four-dimensional scheme).
  }
\end{abstract}

\vspace*{\fill}

 \noindent 
 $^1${\small email address : stefanw@fis.unipr.it}

\newpage


\section{Introduction}
\label{sect:intro}

Higher order computations are a prerequisite for precise predictions
in perturbative QCD.
For many processes next-to-next-to-leading order (NNLO) calculations are required.
This is in particular true for processes involving hadronic jets at high energies.
At the next-to-next-to-leading order level the ingredients for the third order
term in the perturbative expansion for quantities depending on $n$ resolved ``hard'' partons
are the $n$-parton two-loop amplitudes, the $(n+1)$-parton one-loop amplitudes and the
$(n+2)$-parton Born amplitudes.
Taken separately, each one of these contributions is divergent.
The two-loop amplitudes have explicit poles up to $1/\eps^4$, when calculated within
dimensional regularization. As usual the dimensional regularization parameter is
$\eps = 2 -D/2$.
The one-loop amplitudes have explicit poles up to $1/\eps^2$. In addition, integration
over the phase space region where two partons are ``close'' to each other, brings another
two powers of $1/\eps$.
Finally, the Born amplitudes with $(n+2)$ partons are finite, but integration over the phase
space regions, where the $(n+2)$-parton configuration degenerates to a $(n+1)$- or $n$-parton
configuration yields poles in $\eps$ up to $1/\eps^4$.

Due to the large variety of interesting jet observables it is desirable not to perform
a NNLO calculation for a specific observable, but to set up a computer
program, which yields predictions for any infra-red safe observable relevant to the process
under consideration.
This requires to work with fully differential cross-sections and to use 
Monte Carlo techniques for the phase space integration.

The divergences from the separate terms discussed above cancel each other and 
the sum of the three contributions is finite.
However, the cancellation occurs only after the integration over the unresolved phase space
has been performed and prevents thus a naive Monte Carlo approach for a fully exclusive
calculation.
It is therefore necessary to cancel first analytically all infrared divergences and to use
Monte Carlo methods only after this step has been performed.

Due to the universality of the singular behaviour of QCD amplitudes in soft and collinear
limits, the problem can be solved in a process-independent way.
Possible approaches are the extension of the phase space slicing method 
or the subtraction method to NNLO.
Both methods are well understood in NLO calculations.
A third and different approach at NNLO does not try to solve the problem in a process-independent way,
but uses for specific observables the optical theorem to relate phase space integrals
to loop integrals \cite{Anastasiou:2002yz,Anastasiou:2003yy}.
The phase space slicing method has been used for the NLO calculation of the photon + 1 jet rate
in electron-positron annihilation \cite{Gehrmann-DeRidder:1998gf}.
This calculation shares already some features with a full NNLO calculation.

Here, I focus on the subtraction method. This method consists in 
adding and subtracting suitable chosen pieces. The general setup
for the subtraction method at NNLO has been discussed 
in \cite{Weinzierl:2003fx,Moch:2001bi}.
This setup involves three types of subtractions: 
\begin{description}
\item{(a)} Subtraction terms between the $(n+2)$-parton Born amplitudes and the
$(n+1)$-parton one-loop amplitudes. 
These are already known from NLO calculations 
\cite{Frixione:1996ms}-\cite{Catani:2002hc}.
\item{(b)} Subtraction terms between the $(n+1)$-parton one-loop amplitudes and the
$n$-parton two-loop amplitudes. 
\item{(c)} Subtraction terms between the $(n+2)$-parton Born amplitudes and the
$n$-parton two-loop amplitudes. 
\end{description}

In this paper I discuss the subtraction terms corresponding to case (b), e.g.
for one-loop amplitudes with one unresolved parton .
I give all relevant subtraction terms for electron-positron annihilation 
into less than 4 hard partons together
with their integrated counterparts.
The restriction to $e^+ e^- \rightarrow 3 \;\mbox{jets}$ and
$e^+ e^- \rightarrow 2 \;\mbox{jets}$ leads to simplifications, 
since in these processes 
correlations between three hard partons are absent due to colour
conservation \cite{Catani:2000pi}.
Apart from that, $e^+ e^- \rightarrow 3 \;\mbox{jets}$ and $e^+ e^- \rightarrow 2 \;\mbox{jets}$
are from a phenomenological point of view the most relevant processes 
in electron-positron annihilation and the only
ones where the necessary two- and one-loop amplitudes are known
\cite{Garland:2001tf}-\cite{Campbell:1997tv}.
Valuable information for the construction of the subtraction terms for one-loop amplitudes
with one unresolved parton is provided by the known singular behaviour of one-loop
amplitudes in the soft and collinear limits
\cite{Catani:2000pi}, \cite{Bern:1994zx}-\cite{Kosower:2003cz}.

The remaining (and more involved) subtraction terms for double unresolved configurations
will be treated elsewhere.

This paper is organized as follows:
In the next section a short summary on the used notation is given.
Since the subtraction terms are split into several sub-pieces,
this section is included as a ``reference guide'' to improve the readability of the paper.
Sect. \ref{sect:review} reviews the general setup for the subtraction method at NNLO and
the collinear limits of one-loop amplitudes.
In Sect. \ref{squares} the squares and interference terms of the singular factors
are presented.
Sect. \ref{sect:soft} reviews the soft behaviour of one-loop amplitudes and shows
how to obtain the full subtraction terms from the 
known soft and collinear behaviour.
The absence of correlations among three hard partons for processes with less than four
hard partons is also discussed in this section.
In Sect. \ref{sect:subtr} the subtraction terms are given.
The integration over the unresolved phase space is performed in 
Sect. \ref{sect:integrate}.
Finally Sect. \ref{sect:concl} contains a summary and the conclusions.


\section{Notation}
\label{sect:nota}

In this section I give a brief summary for the labelling of the
subtraction terms.
The subtraction terms are classfied accoring to the following
criteria:
\begin{itemize}
\item The number of unresolved particles and the number of loops.
Subtraction terms labelled $d\alpha^{(l,k)}_n$ approximate 
contributions with at most 
$n+k$ partons and at most $l$ loops in the case where 
$k$ partons become unresolved.
In general they approximate not only matrix elements but also lower
order subtraction terms.
To distinguish which term is approximated, the notation
$d\alpha^{(l,k)}_{(l',k') \;n}$ is used and denotes 
an approximation of $d\alpha^{(l',k')}$.
The matrix element $d\sigma^{(l)}$ with $l$ loops is 
identified with $d\alpha^{(l,0)}$.
This paper deals subtraction terms for one-loop amplitudes with
one unresolved parton, e.g. it provides the subtraction term
\bq
d\alpha^{(1,1)}_n.
\eq
This subtraction term has to approximate the interference term of the one-loop
amplitude with the Born amplitude, as well as the integrated subtraction
term $d\alpha^{(0,1)}_{n+1}$.
The approximation for the former is denoted by
$d\alpha^{(1,1)}_{(1,0) \;n}$, the approximation to the later by
$d\alpha^{(1,1)}_{(0,1) \;n}$.
\item The particles involved in the splitting. For $1 \rightarrow 2$
splittings it is sufficient to consider the splittings
$q \rightarrow q g$,
$g \rightarrow g g$ and
$g \rightarrow q \bar{q}$.
\item The decomposition into partial amplitudes. The full amplitude is decomposed
into partial amplitudes with different colour structures. 
The squares and interference terms of the full amplitudes contains therefore
in the off-diagonal entries in colour space products of partial
amplitudes with different cyclic ordering. The notation
\bq
A_{n;i}^\ast A_{n;j}
\eq
is used to denote such a product of 
two partial amplitudes with different cyclic ordering.
\item The decomposition into primitive parts. This follows from the 
decomposition of one-loop amplitudes into primitive amplitudes.
There are three types: A leading-colour piece, denoted by the
subscript ``lc'', a piece proportional to the number of quarks $N_f$
and suppressed by $1/N$, denoted by ``nf'' and a sub-leading colour
piece, suppressed by $1/N^2$ and denoted by ``sc''.
\item The correlations between the hard partons.
In the singular limits of one-loop amplitudes with more than 4 partons
there are in addition to the correlations between two hard partons also
correlations involving three hard partons.
Pieces, which only involve correlations between two partons
(e.g. the emitter and the spectator) are denoted by a subscript ``intr''
(for ``intra'').
The pieces, which involve correlations among three hard partons are classified
whether the emitter is correlated with an additional hard parton 
(denoted by ``emit'') or whether the spectator is correlated with the 
additional hard parton (denoted by ``spec'').
A similar reasoning applies to the singular limit
of the integrated subtraction term $d\alpha^{(0,1)}_{n+1}$.
Here it is useful to distinguish the case ``intr'' further, depending which
two of the three partons from the set 
$\{$ emitter $i$, emitted particle $j$, spectator $k$ $\}$ are correlated.
These cases will be labelled ``intr, i,j'', ``intr, j,k'' and ``intr,i,k''.
\end{itemize}


\section{Review of basic facts and concepts}
\label{sect:review}

This section first reviews the general setup of the subtraction method at NNLO
and discusses then the decomposition of one-loop amplitudes into 
primitive amplitudes.
Throughout this paper I work with renormalized amplitudes. 
Therefore, ultraviolet renormalization is briefly discussed.
Finally, the singular behaviour of one-loop primitive amplitudes in the collinear
limit is summarized.

\subsection{General setup for the subtraction method at NNLO}

The NNLO contribution to an observable ${\cal O}$ whose LO contribution depends on $n$ partons
is given by
\bq
\lefteqn{
\l {\cal O} \r_n^{NNLO} = } \\ 
& &
   \int {\cal O}_{n+2}(p_1,...,p_{n+2}) \; d\sigma_{n+2}^{(0)} 
 + \int {\cal O}_{n+1}(p_1,...,p_{n+1}) \; d\sigma_{n+1}^{(1)} 
 + \int {\cal O}_{n}(p_1,...,p_{n}) \; d\sigma_n^{(2)}. \nonumber 
\eq
The observable ${\cal O}$ has to be infrared safe. 
This requires that whenever a 
$(n+r)$ parton configuration $p_1$,...,$p_{n+r}$ becomes 
kinematically degenerate 
with a $n$ parton configuration $p_1'$,...,$p_{n}'$
we must have
\bq
{\cal O}_{n+r}(p_1,...,p_{n+r}) & \rightarrow & {\cal O}_n(p_1',...,p_n').
\eq
The subscript $n$ for ${\cal O}_{n}$ and $d\sigma_n^{(l)}$
indicates that this contribution is integrated over a $n$-parton phase space.
The various contributions $d\sigma_n^{(l)}$ are given by
\bq
\label{amplcontribut}
d\sigma_{n+2}^{(0)} & = & 
 \left( \left. {\cal A}_{n+2}^{(0)} \right.^\ast {\cal A}_{n+2}^{(0)} \right) 
d\phi_{n+2},  \nonumber \\
d\sigma_{n+1}^{(1)} & = & 
 \left( 
 \left. {\cal A}_{n+1}^{(0)} \right.^\ast {\cal A}_{n+1}^{(1)} 
 + \left. {\cal A}_{n+1}^{(1)} \right.^\ast {\cal A}_{n+1}^{(0)} \right)  
d\phi_{n+1}, \nonumber \\
d\sigma_n^{(2)} & = & 
 \left( 
 \left. {\cal A}_n^{(0)} \right.^\ast {\cal A}_n^{(2)} 
 + \left. {\cal A}_n^{(2)} \right.^\ast {\cal A}_n^{(0)}  
 + \left. {\cal A}_n^{(1)} \right.^\ast {\cal A}_n^{(1)} \right) d\phi_n,
\eq
where ${\cal A}_n^{(l)}$ denotes an amplitude with $n$ external partons and $l$ loops.
$d\phi_n$ is the phase space measure for $n$ partons.
Taken separately, each of the individual contributions 
$d\sigma_{n+2}^{(0)}$, $d\sigma_{n+1}^{(1)}$ and $d\sigma_{n}^{(2)}$  
gives a divergent contribution. Only the sum of all contributions is finite.
To render the individual contributions finite, one adds and subtracts suitable terms \cite{Weinzierl:2003fx}:
\bq
\l {\cal O} \r_n^{NNLO} & = &
 \int \left( {\cal O}_{n+2} \; d\sigma_{n+2}^{(0)} 
             - {\cal O}_{n+1} \circ d\alpha^{(0,1)}_{n+1}
             - {\cal O}_{n} \circ d\alpha^{(0,2)}_{n} 
      \right) \nonumber \\
& &
 + \int \left( {\cal O}_{n+1} \; d\sigma_{n+1}^{(1)} 
               + {\cal O}_{n+1} \circ d\alpha^{(0,1)}_{n+1}
               - {\cal O}_{n} \circ d\alpha^{(1,1)}_{n}
        \right) \nonumber \\
& & 
 + \int \left( {\cal O}_{n} \; d\sigma_n^{(2)} 
               + {\cal O}_{n} \circ d\alpha^{(0,2)}_{n}
               + {\cal O}_{n} \circ d\alpha^{(1,1)}_{n}
        \right).
\eq
Here $d\alpha_{n+1}^{(0,1)}$ is a subtraction term for single unresolved configurations
of Born amplitudes.
This term is already known from NLO calculations.
The term $d\alpha_n^{(0,2)}$ is a subtraction term 
for double unresolved configurations.
Finally, $d\alpha_n^{(1,1)}$ is a subtraction term
for single unresolved configurations involving one-loop amplitudes.
The term $d\alpha_n^{(1,1)}$ is the object of study of this article.

\subsection{Primitive amplitudes}

The squares and the interference terms of the amplitudes in 
eq. (\ref{amplcontribut}) entangle the kinematical structure 
with colour correlations.
For a clear understanding of the factorization properties in soft and collinear limits
it is desirable to disentangle the colour factors from the rest.
This can be done either by introducing colour charge operators \cite{Catani:1997vz}
or by decomposing the full amplitude into partial and primitive amplitudes.
In this paper I follow the second approach.
For Born amplitudes a decomposition into partial amplitudes is sufficient.
However for one-loop amplitudes it is necessary
to decompose the one-loop partial amplitudes further into primitive amplitudes, 
to obtain a clear factorization in singular kinematical limits
\cite{Bern:1999ry}.
Primitive amplitudes can be defined as the sum of all Feynman diagrams 
with a fixed cyclic ordering
of the QCD partons and a definite routing of the external fermion lines through the diagram.
Colour factors and coupling factors are stripped off from primitive amplitudes.
\\
\\
Throughout this paper I use the following normalization for the colour matrices:
\bq
\mbox{Tr} \; T^a T^b = \frac{1}{2} \delta^{ab}.
\eq
Further, $\alpha_s/(2\pi)$ is used as the expansion parameter for the loop expansion, e.g.
a $n$-point amplitude is perturbatively expanded as
\bq
{\cal A}_{n} & = &
 c \; \left( 4 \pi \alpha_s \right)^{n/2-1}
 \left[
        {\cal C} \circ A_{n}^{(0)} 
        + \left( \frac{\alpha_s}{2\pi} \right) {\cal C} \circ A_{n}^{(1)}
                    + \left( \frac{\alpha_s}{2\pi} \right)^2 
                      {\cal C} \circ A_{n}^{(2)}
        + O(\alpha_s^3)
 \right],
\eq
where $c$ is a prefactor containig the electroweak coupling factors 
and ${\cal C}$ is a colour factor. The notation
${\cal C} \circ A_{n}$ indicates that in general this contribution is a sum
of terms
\bq
{\cal C} \circ A^{(l)}_{n} & = &
 \sum {\cal C}^{(l)}_j \cdot A^{(l)}_{n;j}.
\eq
The generic decomposition of one-loop partial amplitudes into primitive
amplitudes is given by
\bq
A_{n}^{(1)} & = & A_{n,lc}^{(1)} + \frac{N_f}{N} A_{n,nf}^{(1)}
                                + \frac{1}{N^2} A_{n,sc}^{(1)},
\eq
where $N$ is the number of colours and $N_f$ the number of quark flavours.
The one-loop singular functions, describing the collinear or soft limit of one-loop
amplitudes, follow the same pattern of decomposition into primitive parts:
\bq
\mbox{Sing}^{(1,1)} & = & \mbox{Sing}_{lc}^{(1,1)} + \frac{N_f}{N} \mbox{Sing}_{nf}^{(1,1)}
                                + \frac{1}{N^2} \mbox{Sing}_{sc}^{(1,1)}.
\eq

\subsection{Ultraviolet renormalization}

The general structure of
the factorization of one-loop primitive amplitudes in singular
(e.g. soft and collinear) limits is
\bq
\label{factsing}
A^{(1)}_{n}
  & = &
  \mbox{Sing}^{(0,1)} 
  \cdot A^{(1)}_{n-1} +
  \mbox{Sing}^{(1,1)} \cdot A^{(0)}_{n-1},
\eq
where the function $\mbox{Sing}$ corresponds either to the eikonal
factor (soft limit) or a splitting function (collinear limit).
Eq. (\ref{factsing}) holds both for bare and renormalized quantitites.
The relation between the renormalized and the bare expression
in the $\overline{\mbox{MS}}$ scheme
for the singular function is
\bq
\mbox{Sing}_{ren}^{(1,1)} & = & S_\eps^{-1} \mu^{2\eps} \; \mbox{Sing}_{bare}^{(1,1)}
                      - \frac{\beta_0}{2 \eps} \; \mbox{Sing}_{bare}^{(0,1)},
\eq
where $\mu^2$ is the renormalization scale,
\begin{eqnarray}
S_\eps & = & \left( 4 \pi \right)^\eps e^{-\eps\gamma_E} \, ,
\end{eqnarray}
is the typical phase-space volume factor in $D =4-2\eps$ dimensions, 
$\gamma_E$ is Euler's constant,
and $\beta_0$ is the first coeffcient of the QCD $\beta$-function:
\begin{eqnarray}
\beta_0 & = & \frac{11}{6} C_A - \frac{2}{3} T_R N_f,
\end{eqnarray}
with the colour factors
\begin{eqnarray}
C_A = N, \;\;\; C_F = \frac{N^2-1}{2N}, \;\;\; T_R = \frac{1}{2}.
\end{eqnarray}

\subsection{Collinear limits}
\label{sectlim}

In the collinear limit tree amplitudes factorize according to
\bq
\label{factcollinearlimit}
A_n^{(0)}(...,p_i,p_j,...) & = & 
 \sum\limits_{\lambda} \; \mbox{Split}^{(0,1)}(p_i,p_j) \; A_{n-1}^{(0)}(...,p,...).
\eq
where $p_i$ and $p_j$ are the momenta of two adjacent legs and
the sum is over all polarizations.
In the collinear limit we parametrize the momenta of the partons $i$ and $j$ as
\cite{Catani:1997vz}
\bq
\label{collinearlimit}
p_i & = & z p + k_\perp - \frac{k_\perp^2}{z} \frac{n}{2 p n }, \nonumber \\
p_j & = & (1-z)  p - k_\perp - \frac{k_\perp^2}{1-z} \frac{n}{2 p n }.
\eq
Here $n$ is a massless four-vector and the transverse component $k_\perp$ satisfies
$2pk_\perp = 2n k_\perp =0$.
The collinear limits occurs for $k_\perp^2 \rightarrow 0$.
The splitting amplitudes $\mbox{Split}^{(0,1)}$ are universal, they depend
only on the two momenta becoming collinear, and not upon the specific amplitude under
consideration.
The splitting functions $\mbox{Split}^{(0,1)}$ are given by
\bq
%
%
%
\mbox{Split}^{(0,1)}_{q \rightarrow q g} & = &
 \frac{1}{s_{ij}} \bar{u}(p_i) \eps\!\!\!/(p_j) u(p),
\nonumber \\
\mbox{Split}^{(0,1)}_{g \rightarrow g g} & = &
 \frac{2}{s_{ij}} \left[
    \eps(p_i) \cdot \eps(p_j) \; p_i \cdot \eps(p)
  + \eps(p_j) \cdot \eps(p) \; p_j \cdot \eps(p_i)
  - \eps(p_i) \cdot \eps(p) \; p_i \cdot \eps(p_j)
 \right],
\nonumber \\
\mbox{Split}^{(0,1)}_{g \rightarrow q \bar{q}} & = &
 \frac{1}{s_{ij}} \bar{u}(p_i) \eps\!\!\!/(p) u(p_j).
\eq
One-loop primitive amplitudes factorize according to
\cite{Bern:1994zx}-\cite{Bern:1999ry}
\bq
\label{factloopcollinearlimit}
\lefteqn{
A_n^{(1)}(...,p_i,p_j,...) = }
 \nonumber \\
 & & 
 \sum\limits_{\lambda} \; \mbox{Split}^{(0,1)}(p_i,p_j) \; A_{n-1}^{(1)}(...,p,...)
 +
 \sum\limits_{\lambda} \; \mbox{Split}^{(1,1)}(p_i,p_j) \; A_{n-1}^{(0)}(...,p,...).
\eq
Here, the splitting amplitudes $\mbox{Split}^{(1,1)}$ occur as a new structure.
The one-loop splitting amplitudes $\mbox{Split}^{(1,1)}$
are decomposed into the primitive parts
\bq
\mbox{Split}^{(1,1)}_{lc},
\;\;\;
\mbox{Split}^{(1,1)}_{nf},
\;\;\;
\mbox{Split}^{(1,1)}_{sc}.
\eq
For the splitting $q \rightarrow q g$ one finds
\cite{Kosower:1999rx,Bern:1999ry}:
%
%
%
\bq
\mbox{Split}^{(1,1)}_{q \rightarrow q g, lc} & = &
  S_\eps^{-1} c_\Gamma \left( \frac{-s_{ij}}{\mu^2} \right)^{-\eps}
     \left\{ \left[ f_1(z)+f_2 \right] \; \mbox{Split}^{(0,1)}_{q \rightarrow q g}
           - f_2 \; \frac{2p_i \cdot \eps(p_j)}{s_{ij}^2} 
                 \bar{u}(p_i) p\!\!\!/_j u(p)
     \right\}
 \nonumber \\
 & &
  - \frac{11}{6\eps} \; \mbox{Split}^{(0,1)}_{q \rightarrow q g},
 \nonumber \\
\mbox{Split}^{(1,1)}_{q \rightarrow q g, nf} & = & 
   \frac{1}{3\eps} \; \mbox{Split}^{(0,1)}_{q \rightarrow q g},
 \nonumber \\
\mbox{Split}^{(1,1)}_{q \rightarrow q g, sc} & = &
  S_\eps^{-1} c_\Gamma \left( \frac{-s_{ij}}{\mu^2} \right)^{-\eps}
     \left\{ \left[ -\frac{1}{\eps^2} - f_1(1-z)+f_2 \right] \; \mbox{Split}^{(0,1)}_{q \rightarrow q g}
 \right. \nonumber \\
 & & \left.
           - f_2 \; \frac{2p_i \cdot \eps(p_j)}{s_{ij}^2} 
                 \bar{u}(p_i) p\!\!\!/_j u(p)
     \right\},
\eq
where the coefficients $f_1(z)$ and $f_2$ are
\bq
f_1(z) & = &
 -\frac{1}{\eps^2} \; {}_2F_1\left(1,-\eps,1-\eps;\frac{-z}{1-z}\right),
 \nonumber \\
f_2 & = & \frac{1-\rho\eps}{2(1-\eps)(1-2\eps)},
\eq
and $c_\Gamma$ is given by
\bq 
c_\Gamma & = & (4\pi)^{\eps} \frac{\Gamma(1+\eps)\Gamma^2(1-\eps)}{\Gamma(1-2\eps)}.
\eq
For the splitting $g \rightarrow g g$ one finds
\bq
\mbox{Split}^{(1,1)}_{g \rightarrow g g, lc} & = &
  S_\eps^{-1} c_\Gamma \left( \frac{-s_{ij}}{\mu^2} \right)^{-\eps}
     \left\{ \left[ \frac{1}{\eps^2} + f_1(z)+ f_1(1-z) \right] \; \mbox{Split}^{(0,1)}_{g \rightarrow g g}
 \right. \nonumber \\
 & & \left. 
           + \frac{1}{2 (3-2\eps)} f_2 \; 
              \frac{2 (p_i-p_j) \cdot \eps(p)}{s_{ij}} 
               \left( 2 \eps(p_i) \cdot \eps(p_j) 
                      - \frac{2 p_i \cdot \eps(p_j) 2 p_j \cdot \eps(p_i)}{s_{ij}}
               \right)
     \right\}
 \nonumber \\
 & &
  - \frac{11}{6\eps} \; \mbox{Split}^{(0,1)}_{g \rightarrow g g},
 \nonumber \\
\mbox{Split}^{(1,1)}_{g \rightarrow g g, nf} & = &
  - S_\eps^{-1} c_\Gamma \left( \frac{-s_{ij}}{\mu^2} \right)^{-\eps}
  \frac{1}{2(1-2\eps)(2-2\eps)(3-2\eps)}
 \nonumber \\
 & & 
              \frac{2 (p_i-p_j) \cdot \eps(p)}{s_{ij}} 
               \left( 2 \eps(p_i) \cdot \eps(p_j) 
                      - \frac{2 p_i \cdot \eps(p_j) 2 p_j \cdot \eps(p_i)}{s_{ij}}
               \right)
   + \frac{1}{3\eps} \; \mbox{Split}^{(0,1)}_{g \rightarrow g g},
 \nonumber \\
\mbox{Split}^{(1,1)}_{g \rightarrow g g, sc} & = & 0.
\eq
Finally, for the splitting $g \rightarrow q \bar{q}$ one finds
\bq
\mbox{Split}^{(1,1)}_{g \rightarrow q \bar{q}, lc} & = &
  S_\eps^{-1} c_\Gamma \left( \frac{-s_{ij}}{\mu^2} \right)^{-\eps}
     \left\{ 
       \left[ \frac{1}{\eps^2} + f_1(z)+ f_1(1-z)
              + \frac{1}{\eps^2} 
                \left(
                      \frac{1}{1-2\eps} - \frac{\eps(1-\rho\eps)}{2(1-\eps)}
 \right. \right. \right.
 \nonumber \\
 & &
 \left. \left. \left.
                      + \frac{2\eps(1-\eps)(1-\rho\eps)}{(1-2\eps)(3-2\eps)}
                \right)
       \right] \; \mbox{Split}^{(0,1)}_{g \rightarrow q \bar{q}}
     \right\}
  - \frac{11}{6\eps} \; \mbox{Split}^{(0,1)}_{q \rightarrow q \bar{q}},
 \nonumber \\
\mbox{Split}^{(1,1)}_{g \rightarrow q \bar{q}, nf} & = &
  - S_\eps^{-1} c_\Gamma \left( \frac{-s_{ij}}{\mu^2} \right)^{-\eps}
              \frac{2(1-\eps) }{\eps (1-2\eps)(3-2\eps)} 
       \; \mbox{Split}^{(0,1)}_{g \rightarrow q \bar{q}}
   + \frac{1}{3\eps} \; \mbox{Split}^{(0,1)}_{q \rightarrow q \bar{q}},
 \nonumber \\
\mbox{Split}^{(1,1)}_{g \rightarrow q \bar{q}, sc} & = &
  S_\eps^{-1} c_\Gamma \left( \frac{-s_{ij}}{\mu^2} \right)^{-\eps}
   \frac{1}{\eps^2}
   \left[
              \frac{1}{1-2\eps}
              - \frac{\eps(1-\rho\eps)}{2(1-\eps)} 
   \right]
       \; \mbox{Split}^{(0,1)}_{g \rightarrow q \bar{q}}.
\eq
The $\eps$-expansion of the hypergeometric function is
\bq
{}_2F_1\left(1,-\eps,1-\eps;\frac{-z}{1-z}\right)
 & = & 1 - \sum\limits_{k=1}^{\infty} \eps^k 
           \; \mbox{Li}_k\left(\frac{-z}{1-z}\right).
\eq 
The hypergeometric function can also be written as
\bq
{}_2F_1\left(1,-\eps,1-\eps;\frac{-z}{1-z}\right)
 & = &
 \Gamma(1+\eps) \Gamma(1-\eps) \left( \frac{z}{1-z} \right)^\eps 
 + 1 - z^\eps \; {}_2F_1\left(\eps,\eps,1+\eps;1-z\right),
 \nonumber \\
\eq
which is useful when studying the soft limit.


\section{Squares and interference terms}
\label{squares}

Due to the sum over spins in eq. (\ref{factcollinearlimit}),
spin correlations are retained in the collinear limit of squared tree amplitudes.
With 
\bq
\label{PLO}
P^{(0,1)} & = &  \sum\limits_{\lambda, \lambda'}
 u(p) \left. \; \mbox{Split}^{(0,1)} \right.^\ast \mbox{Split}^{(0,1)} \;\bar{u}(p)
 \;\;\;\;\mbox{for quarks,}
 \nonumber \\
P^{(0,1)} & = &  \sum\limits_{\lambda, \lambda'}
 \left. \eps^\mu(p) \right.^\ast \;
                \left. \mbox{Split}^{(0,1)} \right.^\ast \mbox{Split}^{(0,1)} \; \eps^\nu(p)
 \;\;\;\;\mbox{for gluons,}
\eq
and the paramterization eq. (\ref{collinearlimit}) one finds
\bq
 P^{(0,1)}_{q \rightarrow q g } & = & 
   \frac{2}{s_{ij}} p\!\!\!/ \left[ \frac{2z}{1-z} + (1 - \rho \eps) (1-z) \right], \nonumber \\
 P^{(0,1)}_{g \rightarrow g g} & = & 
   \frac{2}{s_{ij}}  \left[ - g^{\mu\nu} \left( \frac{2z}{1-z} + \frac{2(1-z)}{z} \right) 
   - 4 (1-\rho \eps) z (1-z) \frac{k^\mu_\perp k^\nu_\perp}{k_\perp^2} \right], \nonumber \\
 P^{(0,1)}_{g \rightarrow q \bar{q}} & = & 
   \frac{2}{s_{ij}} \left[ -g^{\mu\nu} + 4 z (1-z) \frac{k^\mu_\perp k^\nu_\perp}{k_\perp^2} \right].
\eq
I introduced the parameter $\rho$, which specifies the variant of dimensional
regularization:
$\rho  = 1$ for the CDR/HV schemes and $\rho=0$ for the FD scheme.
For the interference term of $\mbox{Split}^{(0,1)}$ with $\mbox{Split}^{(1,1)}$
the analog of eq.(\ref{PLO}) is given by
\bq
P^{(1,1)} & = &  \sum\limits_{\lambda, \lambda'}
 u(p) \left. \; \mbox{Split}^{(0,1)} \right.^\ast \mbox{Split}^{(1,1)} \;\bar{u}(p) \; + c.c.
 \;\;\;\;\mbox{for quarks,}
 \nonumber \\
P^{(1,1)} & = &  \sum\limits_{\lambda, \lambda'}
 \left. \eps^\mu(p) \right.^\ast \;
                \left. \mbox{Split}^{(0,1)} \right.^\ast \mbox{Split}^{(1,1)} \; \eps^\nu(p) \; + c.c.
 \;\;\;\;\mbox{for gluons.}
\eq
Here ``c.c.'' denotes the complex conjugate.
One finds for $q \rightarrow q g$
\bq
\label{oneloopcollinterference}
P^{(1,1)}_{q \rightarrow q g, lc} 
 & = & 
  S_\eps^{-1} c_\Gamma \left( \frac{-s_{ij}}{\mu^2} \right)^{-\eps}
     \left\{ f_1(z) \; P^{(0,1)}_{q \rightarrow q g}
           + f_2 \; \frac{2}{s_{ij}} p\!\!\!/   
                 \left[ 1 - \rho \eps (1-z) \right]
     \right\} 
  - \frac{11}{6\eps} P^{(0,1)}_{q \rightarrow q g} \; + c.c., 
 \nonumber \\
P^{(1,1)}_{q \rightarrow q g, nf} 
 & = & 
  \frac{1}{3\eps} P^{(0,1)}_{q \rightarrow q g} \; + c.c., 
 \nonumber \\
P^{(1,1)}_{q \rightarrow q g, sc} 
 & = & 
  S_\eps^{-1} c_\Gamma \left( \frac{-s_{ij}}{\mu^2} \right)^{-\eps}
     \left\{ \left[ -\frac{1}{\eps^2} - f_1(1-z) \right] \; P^{(0,1)}_{q \rightarrow q g}
           + f_2 \; \frac{2}{s_{ij}} p\!\!\!/   
                 \left[ 1 - \rho \eps (1-z) \right]
     \right\} 
  + c.c..
 \nonumber \\
\eq
For the splitting $g \rightarrow g g$ one has
\bq
P^{(1,1)}_{g \rightarrow g g, lc} 
 & = & 
  S_\eps^{-1} c_\Gamma \left( \frac{-s_{ij}}{\mu^2} \right)^{-\eps}
     \left\{ \left[ \frac{1}{\eps^2} + f_1(z)+ f_1(1-z) \right] \; P^{(0,1)}_{g \rightarrow g g}
 \right. \nonumber \\
 & & \left. 
           - \frac{1}{3-2\eps} f_2 \; 
             \frac{2}{s_{ij}}
             \left[ 4 - 8 \rho \eps z (1-z) \right]
             \frac{k^\mu_\perp k^\nu_\perp}{k_\perp^2}
     \right\}
  - \frac{11}{6\eps} \; P^{(0,1)}_{g \rightarrow g g} \; + c.c.,
 \nonumber \\
P^{(1,1)}_{g \rightarrow g g, nf} & = &
  S_\eps^{-1} c_\Gamma \left( \frac{-s_{ij}}{\mu^2} \right)^{-\eps}
  \frac{1}{(1-2\eps)(2-2\eps)(3-2\eps)}
             \frac{2}{s_{ij}}
             \left[ 4 - 8 \rho \eps z (1-z) \right]
             \frac{k^\mu_\perp k^\nu_\perp}{k_\perp^2}
 \nonumber \\
 & & 
   + \frac{1}{3\eps} \; P^{(0,1)}_{g \rightarrow g g} \; + c.c.,
 \nonumber \\
P^{(1,1)}_{g \rightarrow g g, sc} & = & 0.
\eq
Finally for the splitting $g \rightarrow q \bar{q}$ one finds
\bq
P^{(1,1)}_{g \rightarrow q \bar{q}, lc} & = &
  S_\eps^{-1} c_\Gamma \left( \frac{-s_{ij}}{\mu^2} \right)^{-\eps}
       \left[ \frac{1}{\eps^2} + f_1(z)+ f_1(1-z)
              + \frac{1}{\eps^2} 
                \left(
                      \frac{1}{1-2\eps} - \frac{\eps(1-\rho\eps)}{2(1-\eps)}
 \right. \right. 
 \nonumber \\
 & &
 \left. \left. 
                      + \frac{2\eps(1-\eps)(1-\rho\eps)}{(1-2\eps)(3-2\eps)}
                \right)
       \right] \; P^{(0,1)}_{g \rightarrow q \bar{q}}
  - \frac{11}{6\eps} \; P^{(0,1)}_{q \rightarrow q \bar{q}} \; + c.c.,
 \nonumber \\
P^{(1,1)}_{g \rightarrow q \bar{q}, nf} & = &
  - S_\eps^{-1} c_\Gamma \left( \frac{-s_{ij}}{\mu^2} \right)^{-\eps}
              \frac{2(1-\eps) }{\eps (1-2\eps)(3-2\eps)} 
       \; P^{(0,1)}_{g \rightarrow q \bar{q}}
   + \frac{1}{3\eps} \; P^{(0,1)}_{q \rightarrow q \bar{q}} \; + c.c.,
 \nonumber \\
P^{(1,1)}_{g \rightarrow q \bar{q}, sc} & = &
  S_\eps^{-1} c_\Gamma \left( \frac{-s_{ij}}{\mu^2} \right)^{-\eps}
   \frac{1}{\eps^2}
   \left[
              \frac{1}{1-2\eps}
              - \frac{\eps(1-\rho\eps)}{2(1-\eps)} 
   \right]
       \; P^{(0,1)}_{g \rightarrow q \bar{q}} \; + c.c..
\eq

\section{Soft gluons and correlations}
\label{sect:soft}

If a single gluon becomes soft, the partial tree amplitude factorizes according to
\bq
A_n^{(0)}(...,p_i,p_j,p_k,...) & = & \mbox{Eik}^{(0,1)}(p_i,p_j,p_k) A_{n-1}^{(0)}(...,p_i,p_k,...),
\eq
where the eikonal factor is given by
\bq
\mbox{Eik}^{(0,1)}(p_i,p_j,p_k)
 & = & 
 \frac{2 p_i \cdot \eps(p_j)}{s_{ij}} - \frac{2 p_k \cdot \eps(p_j)}{s_{jk}}.
\eq
In the soft-gluon limit,
a one-loop primitive amplitude factorizes according to
\bq
A^{(1)}_{n}(...,p_i,p_j,p_k,...)
  & = &
  \mbox{Eik}^{(0,1)}(p_i,p_j,p_k) 
  \cdot A^{(1)}_{n-1}(...,p_i,p_k,...) \nonumber \\
& &
+
  \mbox{Eik}^{(1,1)}(p_i,p_j,p_k) \cdot A^{(0)}_{n-1}(...,p_i,p_k,...).
\eq
The one-loop eikonal functions are given by
\cite{Bern:1999ry}
\bq
\mbox{Eik}^{(1,1)}_{lc}(p_i,p_j,p_k)
 & = &
 \left[
 - \frac{S_\eps^{-1} c_\Gamma}{\eps^2} 
 \Gamma(1+\eps) \Gamma(1-\eps)
 \left( \frac{\mu^2 (-s_{ik})}{(-s_{ij}) (-s_{jk}) } \right)^\eps
 \right. 
 \nonumber \\
 & & \left.
 - \frac{11}{6\eps} 
 \right]
 \mbox{Eik}^{(0,1)}(p_i,p_j,p_k),
 \nonumber \\
\mbox{Eik}^{(1,1)}_{nf}(p_i,p_j,p_k)
 & = &
 \frac{1}{3\eps} 
 \mbox{Eik}^{(0,1)}(p_i,p_j,p_k),
 \nonumber \\
\mbox{Eik}^{(1,1)}_{sc}(p_i,p_j,p_k)
 & = & 0.
\eq
As in the tree-level case with one soft gluon, the eikonal factor
is independent of the variant of the dimensional regularization scheme used.
The square of the Born eikonal factor is given by
\bq
\label{sqreikonal}
\sum\limits_{\lambda_j}
\left. \; \mbox{Eik}^{(0,1)}(p_i,p_j,p_k) \right.^\ast 
       \mbox{Eik}^{(0,1)}(p_i,p_j,p_k) & = & 
 4 \frac{s_{ik}}{s_{ij}s_{jk}}.
\eq
For sub-leading colour contributions we have in addition the case of interference terms
of two primitive amplitudes with non-identical cyclic ordering of the external partons.
Therefore we need
\bq
\label{infeikonal}
\sum\limits_{\lambda_j}
\left. \; \mbox{Eik}^{(0,1)}(p_i,p_j,p_l) \right.^\ast 
          \mbox{Eik}^{(0,1)}(p_i,p_j,p_k) & = & 
 2 \frac{s_{ik}}{s_{ij}s_{jk}}
 + 2 \frac{s_{il}}{s_{ij}s_{jl}}
 - 2 \frac{s_{kl}}{s_{kj}s_{jl}},
 \nonumber \\
\sum\limits_{\lambda_j}
\left. \; \mbox{Eik}^{(0,1)}(p_h,p_j,p_l) \right.^\ast 
          \mbox{Eik}^{(0,1)}(p_i,p_j,p_k) & = & 
 2 \frac{s_{il}}{s_{ij}s_{jl}}
 + 2 \frac{s_{hk}}{s_{hj}s_{jk}}
 - 2 \frac{s_{kl}}{s_{kj}s_{jl}}
 - 2 \frac{s_{hi}}{s_{hj}s_{ji}}.
\eq
It should be noted that the product of the two eikonal factors, which share only one or no hard
parton decomposes into squares of eikonal factors.
From eq. (\ref{sqreikonal}) and eq. (\ref{infeikonal}) one obtains at NLO all
subtraction terms by partial fractioning
\bq
\label{parfraceik}
 \frac{s_{ik}}{s_{ij}s_{jk}}
  & = &
 \frac{s_{ik}}{s_{ij}(s_{ij}+s_{jk})} 
+
 \frac{s_{ik}}{s_{jk}(s_{ij}+s_{jk})} 
\eq
and by extending then each term on the r.h.s of eq. (\ref{parfraceik}) to the full
dipole splitting function
\cite{Weinzierl:1998ki}.
For one-loop amplitudes with one unresolved parton, additional $\eps$-powers of eikonal factors
occur.
For example
\bq
\lefteqn{
\left( \frac{\mu^2 (-s_{ik})}{(-s_{ij}) (-s_{jk}) } \right)^\eps
\sum\limits_{\lambda_j}
\left. \; \mbox{Eik}^{(0,1)}(p_i,p_j,p_l) \right.^\ast 
          \mbox{Eik}^{(0,1)}(p_i,p_j,p_k) 
 = }
 \nonumber \\
 & & 
2 \left( \frac{\mu^2 (-s_{ik})}{(-s_{ij}) (-s_{jk}) } \right)^\eps
 \left[
 \frac{s_{ik}}{s_{ij}(s_{ij}+s_{jk})} + \frac{s_{il}}{s_{ij}(s_{ij}+s_{jl})}
 + \frac{s_{ik}}{s_{jk}(s_{ij}+s_{jk})} - \frac{s_{kl}}{s_{jk}(s_{jk}+s_{jl})} 
 \right. \nonumber \\
 & & \left.
 + \frac{s_{il}}{s_{jl}(s_{ij}+s_{jl})} - \frac{s_{kl}}{s_{jl}(s_{jk}+s_{jl})} 
 \right].
\eq
Here one can distinguish three types, depending on the hard partons which
occur in the $\eps$-power of the eikonal factor.
Contributions of the form
\bq
\left( \frac{\mu^2 (-s_{ik})}{(-s_{ij}) (-s_{jk}) } \right)^\eps
\frac{s_{ik}}{s_{ij}(s_{ij}+s_{jk})}
\eq
involve in addition to the emitted soft parton $j$ only the emitter $i$ and the spectator $k$.
These contributions are labelled ``intr''.
Terms of the form 
\bq
\left( \frac{\mu^2 (-s_{il})}{(-s_{ij}) (-s_{jl}) } \right)^\eps
\frac{s_{ik}}{s_{ij}(s_{ij}+s_{jk})}
\eq
are labelled ``emit'', since the eikonal factor which occurs to the $\eps$-power
corresponds to the emission of a soft parton $j$ from an antenna formed by the emitter
$i$ and an additional hard parton $l$.
Finally, terms of the form
\bq
\left( \frac{\mu^2 (-s_{kl})}{(-s_{kj}) (-s_{jl}) } \right)^\eps
\frac{s_{ik}}{s_{ij}(s_{ij}+s_{jk})}
\eq
are labelled ``spec'', since the eikonal factor which occurs to the $\eps$-power
corresponds to the emission of a soft parton $j$ from an antenna formed by the spectator
$k$ and an additional hard parton $l$.

In general, these correlations with an additional hard parton $l$ have to be taken into
account.
However, for the processes $e^+ e^- \rightarrow 2 \;\mbox{jets}$ and
$e^+ e^- \rightarrow 3 \;\mbox{jets}$ these correlations are absent.
This is obvious for the NNLO calculation
$e^+ e^- \rightarrow 2 \;\mbox{jets}$, since there are only two hard partons.
It also holds for the NNLO calculation $e^+ e^- \rightarrow 3 \;\mbox{jets}$
due to colour conservation.
The vanishing of the accompanying colour factor is most easily seen when working
within the formalism of colour-charge operators.
In the formalism of colour charge operators it is convenient to introduce
for an amplitude with $n+1$ partons
an orthogonal basis of unit vectors ${|c_1,c_2,...,c_{n+1}\r}$ 
in the colour space.
An abstract vector $|{\cal M}_{n+1}\r $ in this colour space is then defined as the
projection of the amplitude ${\cal A}_{n+1}$ onto this basis: 
\bq
  {\cal A}_{n+1}   & = & \l c_1,c_2,...,c_{n+1} | {\cal M}_{n+1} \r.
\eq
The vector $|{\cal M}_{n+1}\r $ is expanded 
in the coupling $\alpha_s$ as
\bq
  | {\cal M}_{n+1} \r & = & 
 c \; \left( 4 \pi \alpha_s \right)^{(n-1)/2} S_\eps^{-(n-1)/2}
 \left[
 \left| {\cal M}^{(0)}_{n+1} \right\r
 + \left( \frac{\alpha_s}{2\pi} \right) \left| {\cal M}^{(1)}_{n+1} \right\r
 + {\cal O}(\alpha_s^2)
 \right].
\eq
In the soft limit where the momentum of parton $j$ becomes soft, 
the $n+1$-parton amplitudes behave as
\bq
\label{catanisoft}
\left| {\cal M}^{(0)}_{n+1} \right\r
 & = & \sqrt{4 \pi \alpha_s} \; S_\eps^{-1/2} \eps^\mu(p_j) \; {\bf J}^{(0)}_\mu(p_j)
       \left| {\cal M}^{(0)}_{n} \right\r,
 \nonumber \\
\left| {\cal M}^{(1)}_{n+1} \right\r
 & = & \sqrt{4 \pi \alpha_s} \; S_\eps^{-1/2} \eps^\mu(p_j) 
       \left[
         {\bf J}^{(0)}_\mu(p_j) \left| {\cal M}^{(1)}_{n} \right\r
         +\left( \frac{\alpha_s}{2\pi} \right)
          {\bf J}^{(1)}_\mu(p_j) \left| {\cal M}^{(0)}_{n} \right\r
       \right],
\eq
where the soft currents are given by
\cite{Catani:2000pi}
\bq
\label{softcurrent}
\lefteqn{
{\bf J}^{a\;(0)}_\mu(p_j)
 = 
 \sum\limits_i {\bf T}^a_i \frac{p_i^\mu}{p_i \cdot p_j},
}
 \nonumber \\
\lefteqn{
{\bf J}^{a\;(1)}_\mu(p_j)
 = }
 \nonumber \\
 & & 
 - \frac{S_\eps^{-1} c_\Gamma}{\eps^2} 
 \Gamma(1+\eps) \Gamma(1-\eps)
 \frac{1}{2} i f^{abc} 
 \sum\limits_{i \neq k}
  {\bf T}^b_i {\bf T}^c_k
  \left( \frac{\mu^2 (-s_{ik})}{(-s_{ij}) (-s_{jk}) } \right)^\eps
  \left(
          \frac{p_i^\mu}{p_i \cdot p_j} 
        - \frac{p_k^\mu}{p_k \cdot p_j}
  \right) 
 \nonumber \\
 & &
 - \frac{\beta_0}{2\eps} {\bf J}^{a\;(0)}_\mu(p_j).
\eq 
The colour charge operators ${\bf T}_i$ for a quark, gluon and antiquark in the final state are
\bq
\mbox{quark :} & & \l ... q_i ... | T_{ij}^a | ... q_j ... \r, \nonumber \\
\mbox{gluon :} & & \l ... g^c ... | i f^{cab} | ... g^b ... \r, \nonumber \\
\mbox{antiquark :} & & \l ... \bar{q}_i ... | \left( - T_{ji}^a \right) | ... \bar{q}_j ... \r.
\eq
Colour conservation reads
\bq 
\sum\limits_{i} {\bf T}_i \left| {\cal M}^{(0)}_{n} \right\r & = & 0.
\eq
From eq. (\ref{softcurrent}) it follows that terms labelled ``spec'' or 
``emit'' are accompagnied by a colour factor
\bq
i f^{abc} \; {\bf T}^a_i {\bf T}^b_k {\bf T}^c_l.
\eq
This colour factor equals zero for processes with less than four hard partons.
Therefore for the major applications $e^+ e^- \rightarrow \;\mbox{3 jets}$
and $e^+ e^- \rightarrow \;\mbox{2 jets}$
these correlations are absent.

Similar considerations apply to the approximation $d\alpha^{(1,1)}_{(0,1)\;n}$
of the integrated subtraction term $d\alpha^{(0,1)}_{n+1}$.
The integrated subtraction term $d\alpha^{(0,1)}_{n+1}$ can be written
as
\bq
d\alpha^{(0,1)}_{n+1} & = & 
 \frac{c}{8 \pi^2} 
\left\l {\cal M}^{(0)}_{n+1} \right|
{\bf I}
\left| {\cal M}^{(0)}_{n+1} \right\r,
\eq
with the insertion operator
\bq
{\bf I} = - \sum\limits_{l \neq m}
 \frac{{\bf T}^a_l {\bf T}^a_m}{{\bf T}^2_l} {\cal I}_{lm},
&&
{\cal I}_{lm} =  
 \frac{S_\eps^{-1} \left( 4 \pi \right)^\eps}{\Gamma(1-\eps)}
  \left( \frac{s_{lm}}{\mu^2} \right)^{-\eps}
 {\cal V}_l.
\eq
The singular factors ${\cal V}_l$ are given for a quark by
\bq
{\cal V}_q & = & C_F {\cal V}^{(0,1)}_{q \rightarrow q g},
\eq
and for a gluon by
\bq
{\cal V}_g & = & C_A {\cal V}^{(0,1)}_{g \rightarrow g g}
 + N_f {\cal V}^{(0,1)}_{g \rightarrow q \bar{q}}.
\eq
The functions ${\cal V}^{(0,1)}$ will be given 
in eq. (\ref{ressingunresolved}).
In the soft limit the amplitude $\left| {\cal M}^{(0)}_{n+1} \right\r$ 
factorizes
as in eq. (\ref{catanisoft}).
If we denote the emitter by $i$, the emitted soft parton by $j$ and
the spectator by $k$, we can distinguish the following cases:
\begin{itemize}
\item $l,m \in \{i,j,k\}$. In this case we obtain terms of the form
\bq
\left( \frac{s_{ij}}{\mu^2} \right)^{-\eps}
\frac{s_{ik}}{s_{ij}(s_{ij}+s_{jk})},
 \;\;\;
\left( \frac{s_{jk}}{\mu^2} \right)^{-\eps}
\frac{s_{ik}}{s_{ij}(s_{ij}+s_{jk})},
 \;\;\;
\left( \frac{s_{ik}}{\mu^2} \right)^{-\eps}
\frac{s_{ik}}{s_{ij}(s_{ij}+s_{jk})}.
\eq
These cases are labelled ``intr,$\;i,j$'', ``intr,$\;j,k$'' and ``intr,$\;i,k$'',
respectively. 
\item Either $l$ or $m$ in $\{i,j,k\}$, and the other one distinct.
In this case one obtains terms of the form
\bq
\label{approxemitspec}
\left( \frac{s_{il}}{\mu^2} \right)^{-\eps}
\frac{s_{ik}}{s_{ij}(s_{ij}+s_{jk})},
 \;\;\;
\left( \frac{s_{kl}}{\mu^2} \right)^{-\eps}
\frac{s_{ik}}{s_{ij}(s_{ij}+s_{jk})},
 \;\;\;
\left( \frac{s_{jl}}{\mu^2} \right)^{-\eps}
\frac{s_{ik}}{s_{ij}(s_{ij}+s_{jk})}.
\eq
With an argument similar to the one used for the soft behaviour
of one-loop amplitudes, one can show that the colour factor for
the case $s_{jl}^{-\eps}$ vanishes for amplitudes with less than
four hard partons.
Therefore only the first two cases have to be considered.
They will be labelled ``emit'' and ``spec''.
\item $l,m \notin \{i,j,k\}$. This case requires at least four hard
partons and is therefore not relevant to 
$e^+ e^- \rightarrow \;\mbox{3 jets}$
and $e^+ e^- \rightarrow \;\mbox{2 jets}$.
\end{itemize}


\section{Subtraction terms}
\label{sect:subtr}

The approximation terms $d\alpha_n^{(1,1)}$ and  
$d \alpha^{(0,1)}_{n+1}$ are written as a sum over dipoles
\cite{Weinzierl:2003fx,Catani:1997vz}:
\bq
d \alpha^{(1,1)}_{n} =  
 \sum\limits_{\mbox{\tiny topologies $T$}} {\cal D}_{n}^{(1,1)}(T),
 & &
d \alpha^{(0,1)}_{n+1} =  
 \sum\limits_{\mbox{\tiny topologies $T$}} {\cal D}_{n+1}^{(0,1)}(T).
\eq
The splitting topologies are here just $1 \rightarrow 2$ splittings.
A dipol is constructed from amputated amplitudes, evaluated with mapped momenta
and dipole splitting functions.
By an amputated amplitude I mean an amplitude where a polarization vector 
(or external spinor for fermions) has been removed.
I denote amputated amplitudes by $\left| A \right\r$.
If the amputated parton is a gluon, one has
\bq
\left| A_n(...,p,...) \right\r
 & = & 
 \frac{\partial}{\partial \eps_\mu(p)} A_n(...,p,...). 
\eq
If the amputated parton is a quark, one has
\bq
\left| A_n(...,p,...) \right\r
 & = & 
 \frac{\partial}{\partial \bar{u}(p)} A_n(...,p,...).
\eq
The dipole subtraction terms are obtained by sandwiching a dipole splitting function
in between these amputated amplitudes:
\bq
{\cal D}_{n+1}^{(0,1)}(T) & = &
 \left| c \right|^2 \left( 4 \pi \alpha_s \right)^{n} {\cal C}_{ij} \; \left\l A_{n+1;i}^{(0)}(...,p_e,...) \right|
 {\cal P}^{(0,1)}(T)
 \left| A_{n+1;j}^{(0)}(...,p_e,...) \right\r d\phi_{n+2}.
\eq
Here, $c$ denotes an overall prefactor and ${\cal C}$ is a colour factor. 
The subtraction term $d\alpha_n^{(1,1)}$
approximates
\bq
\lefteqn{
d\sigma_{n+1}^{(1)} + d\alpha_{n+1}^{(0,1)}
 = }
 \nonumber \\
 & & 
 \sum
 \frac{\left| c \right|^2 \left(4 \pi \alpha_s \right)^{n}}{8 \pi^2} \; {\cal C}_{ij}
 \left[
         \left. A_{n+1;i}^{(0)} \right.^\ast A_{n+1;j}^{(1)}
       + \left. A_{n+1;i}^{(1)} \right.^\ast A_{n+1;j}^{(0)}
       + \left. A_{n+1;i}^{(0)} \right.^\ast {\cal I}_{ij} A_{n+1;j}^{(0)}
 \right]
 d\phi_{n+1}.
\eq 
${\cal I}_{ij}$ is obtained from integrating $d\alpha_{n+1}^{(0,1)}$ 
over the unresolved phase space.
The subtraction terms for this contribution are written as
\bq
\label{oneloopsubtrterms}
{\cal D}_{n}^{(1,1)}(T) & = &
 \frac{\left| c \right|^2 \left(4 \pi \alpha_s \right)^{n}}{8\pi^2} \; {\cal C}_{ij}
  \left[
         \left\l A_{n;i}^{(0)} \right| {\cal P}^{(0,1)}(T) \left| A_{n;j}^{(1)} \right\r 
       + \left\l A_{n;i}^{(1)} \right| {\cal P}^{(0,1)}(T) \left| A_{n;j}^{(0)} \right\r 
 \right. \nonumber \\
 & & \left.
       + \left\l A_{n;i}^{(0)} \right| {\cal P}^{(1,1)}_{(1,0)}(T) \left| A_{n;j}^{(0)} \right\r 
       + \left\l A_{n;i}^{(0)} \right| \left. {\cal P}^{(1,1)}_{(1,0)}(T) \right.^\ast \left| A_{n;j}^{(0)} \right\r 
 \right. \nonumber \\
 & & \left.
       + \left\l A_{n;i}^{(0)} \right| {\cal P}^{(1,1)}_{(0,1)}(T) \left| A_{n;j}^{(0)} \right\r 
  \right] d\phi_{n+1}.
\eq
Here, in the first and second term 
the NLO splitting function ${\cal P}^{(0,1)}$ is sandwiched between
$A_{n}^{(0)}$ and $A_{n}^{(1)}$.
If we recall the factorization formula eq. (\ref{factsing}) for one-loop amplitudes
\bq
A^{(1)}_{n+1}
  & = &
  \mbox{Sing}^{(0)} 
  \cdot A^{(1)}_{n} +
  \mbox{Sing}^{(1)} \cdot A^{(0)}_{n},
\eq
then the first two terms of eq. (\ref{oneloopsubtrterms}) approximate exactly the singular part 
proportional to $\mbox{Sing}^{(0)}$.
The third and fourth term in eq. (\ref{oneloopsubtrterms}), which involve the splitting function
${\cal P}^{(1,1)}_{(1,0)}$, approximate the singular part 
proportional to $\mbox{Sing}^{(1)}$.
Finally, ${\cal P}^{(1,1)}_{(0,1)}$ in the last term of eq. (\ref{oneloopsubtrterms})
is an approximation to $d\alpha_{n+1}^{(0,1)}$.
\\
The amputated amplitudes are evaluated with mapped momenta.
The mapping of momenta relates a $n+1$ parton configuration to
a $n$ parton configuration (or a $n+2$ parton configuration to
a $n+1$ parton configuration).
Such a mapping has to satisfy momentum conservation and the on-mass-shell conditions.
Furthermore it must have the right behaviour in the singular limits.
Several choices for such a mapping exist \cite{Catani:1997vz,Kosower:1998zr}.
One possible choice relates three parton momenta of the $n+1$ parton configuration
to two parton momenta of the $n$ parton configuration, while leaving
the remaining $n-2$ parton momenta unchanged.
This mapping is given by
\bq
\label{reconstructCataniSeymour}
p_e & = & p_i + p_j - \frac{y}{1-y} p_k, 
\nonumber \\
p_s & = & \frac{1}{1-y} p_k.
\eq
$p_i$, $p_j$ and $p_k$ are the momenta of the $n+1$ parton configuration,
and $p_e$ and $p_s$ are the resulting momenta of the of the $n$ parton configuration.
The variables $y$ and $z$ are given by
\bq
y = \frac{s_{ij}}{s_{ijk}}, & & z = \frac{s_{ik}}{s_{ik}+s_{jk}}.
\eq
It is also useful to introduce the vector $k_\perp$, defined by
\bq
k_\perp & = & (1-z) p_i - z p_j - \left( 1 -2 z \right) \frac{y}{1-y} p_k.
\eq
The dipole splitting functions ${\cal P}^{(0,1)}$ read
\bq
\label{nlosplittingfcts}
{\cal P}^{(0,1)}_{q \rightarrow q g} & = & 
   \frac{2}{s_{ijk}} \frac{1}{y} p\!\!\!/_{e} 
         \left[ \frac{2}{1-z(1-y)} - 2  + \left( 1 - \rho \eps \right) (1-y) (1-z) \right],
\nonumber \\
{\cal P}^{(0,1)}_{g \rightarrow g g} & = & 
   \frac{2}{s_{ijk}} \frac{1}{y}  
         \left[ - g^{\mu\nu} \left( \frac{2}{1-z(1-y)} - 2 \right)
           + \left( 1 - \rho \eps \right) \frac{4 r (1-y)^2}{y s_{ijk}} k_\perp^\mu k_\perp^\nu \right],
\nonumber \\
{\cal P}^{(0,1)}_{g \rightarrow q \bar{q}} & = & 
   \frac{2}{s_{ijk}} \frac{1}{y}  
         \left[ - \frac{1}{2} g^{\mu\nu} 
           - \frac{4 r (1-y)^2}{y s_{ijk}} k_\perp^\mu k_\perp^\nu \right].
\eq
For the $q \rightarrow q g$ splitting, the quark polarization
matrix $p\!\!\!/_{e}$ is included in the definition of the dipole splitting function.
For the $g \rightarrow g g$ and $g \rightarrow q \bar{q}$ splittings,
the full splitting function is shared between the dipoles with
spectators $k$ and $k'$, where $k$ and $k'$ are the partons adjacent
to the $(i,j)$ pair in the colour ordered 
amplitude $A^{(0)}(...,k',i,j,k,...)$.
For the $g \rightarrow g g$ splitting the dipole splitting function
is chosen in such a way that the singularity when parton $j$ becomes
soft resides in ${\cal P}_{g \rightarrow g g}(p_i,p_j;p_k)$, where as the singularity when parton $i$
becomes soft resides in ${\cal P}_{g \rightarrow g g}(p_j,p_i;p_{k'})$.
The parameter $r$ is either chosen as 
\bq
r & = & \frac{1}{2},
\eq
or as 
\bq
\label{choice2r}
r & = & \frac{2p_{k'}p_e}{2p_{k'}p_e+2p_e p_s},
\eq
where $k'$ is the other parton adjacent to the emitter $i$ in one of the amplitudes.
The $g \rightarrow g g$ and $g \rightarrow q \bar{q}$ splittings also
involve spin correlation through the spin correlation tensor $k_\perp^\mu k_\perp^\nu$.
Note that the spin correlation tensor $k_\perp^\mu k_\perp^\nu$
is orthogonal to $p_e$ and $p_s$:
\bq
2 p_e k_\perp = 2 p_s k_\perp = 0.
\eq
Further note that the contraction of $p_e$ into an amputated amplitude vanishes
due to gauge invariance.
\\
The parameter $\rho$ specifies the variant of dimensional regularization:
$\rho = 1$ corresponds to the CDR/HV schemes and $\rho=0$ to a four-dimensional
scheme.
\\
The splitting functions ${\cal P}^{(1,1)}_{(1,0)}$ for the splitting $q \rightarrow q g$
are given by
\bq
\lefteqn{
{\cal P}^{(1,1)}_{(1,0)\; q \rightarrow q g, lc, corr} 
 =  
 S_\eps^{-1} c_\Gamma \left( \frac{-s_{ijk}}{\mu^2} \right)^{-\eps} 
  y^{-\eps}
 }
 \nonumber \\
 & &
 \times
     \left\{ 
         g_{1, corr}(y,z) \; {\cal P}^{(0,1)}_{q \rightarrow q g}
         + f_2 \frac{2}{s_{ijk}} \frac{1}{y} p\!\!\!/_{e} 
               \left[ 1 - \rho \eps (1-y) (1-z) \right]
     \right\} 
     - \frac{11}{6\eps} {\cal P}^{(0,1)}_{q \rightarrow q g},
 \nonumber \\
\lefteqn{
{\cal P}^{(1,1)}_{(1,0)\; q \rightarrow q g, nf} 
 =  
  \frac{1}{3\eps} {\cal P}^{(0,1)}_{q \rightarrow q g}, 
 }
 \nonumber \\
\lefteqn{
{\cal P}^{(1,1)}_{(1,0)\; q \rightarrow q g, sc} 
 =  
  S_\eps^{-1} c_\Gamma \left( \frac{-s_{ijk}}{\mu^2} \right)^{-\eps}
    y^{-\eps}
 }
 \nonumber \\
 & &
 \times
     \left\{ \left[ -\frac{1}{\eps^2} - g_{1, intr}(y,1-z) \right] 
                    \; {\cal P}^{(0,1)}_{q \rightarrow q g}
           + f_2 \; \frac{2}{s_{ijk}} \frac{1}{y} p\!\!\!/_{e}   
                 \left[ 1 - \rho \eps (1-y) (1-z) \right]
     \right\}.
\eq
The label ``$corr$'' takes in general the values $\{intr,emit,spec\}$.
However for $e^+ e^- \rightarrow 3 \;\mbox{jets}$ or 
$e^+ e^- \rightarrow 2 \;\mbox{jets}$ only $g_{1, intr}$ is relevant.
The function $g_{1, intr}$ is given by
\bq
\lefteqn{
g_{1, intr}(y,z) 
 = }
 \nonumber \\
 & &
  - \frac{1}{\eps^2} 
 \left[ \Gamma(1+\eps) \Gamma(1-\eps) \left( \frac{z}{1-z} \right)^\eps 
        + 1 
        - (1-y)^\eps z^\eps \; {}_2F_1\left( \eps, \eps, 1+\eps; (1-y)(1-z) \right) \right].
 \nonumber \\
\eq
For the numerical integration over the $(n+1)$-parton phase space we need the 
$\eps$-expansion of the hypergeometric function ${}_2F_1(\eps,\eps,1+\eps;x)$:
\bq
{}_2F_1(\eps,\eps,1+\eps;x) & = &
 1 + \eps^2 \; \mbox{Li}_2(x) 
 + {\cal O}(\eps^3).
\eq
For the splitting $g \rightarrow g g$ one has
\bq
{\cal P}^{(1,1)}_{(1,0)\; g \rightarrow g g, lc, corr}
 & = & 
  S_\eps^{-1} c_\Gamma \left( \frac{-s_{ijk}}{\mu^2} \right)^{-\eps}
   y^{-\eps}
     \left\{ \left[ \frac{1}{\eps^2} + g_{1,corr}(y,z)+ g_{1,intr}(y,1-z) \right] \; {\cal P}^{(0,1)}_{g \rightarrow g g}
 \right. \nonumber \\
 & & \left. 
           + \frac{1}{3-2\eps} f_2 \; 
             \frac{2}{s_{ijk}^2} \frac{r}{y^2}
             \left[ \frac{4}{z(1-z)} - 8 \rho \eps (1-y)^2 \right]
             k^\mu_\perp k^\nu_\perp
     \right\}
  - \frac{11}{6\eps} \; {\cal P}^{(0,1)}_{g \rightarrow g g},
 \nonumber \\
{\cal P}^{(1,1)}_{(1,0)\; g \rightarrow g g, nf}
 & = &
  - S_\eps^{-1} c_\Gamma \left( \frac{-s_{ijk}}{\mu^2} \right)^{-\eps}
   y^{-\eps}
  \frac{1}{(1-2\eps)(2-2\eps)(3-2\eps)}
 \nonumber \\
 & & 
 \times
             \frac{2}{s_{ijk}^2} \frac{r}{y^2}
             \left[ \frac{4}{z(1-z)} - 8 \rho \eps (1-y)^2\right]
             k^\mu_\perp k^\nu_\perp
   + \frac{1}{3\eps} \; {\cal P}^{(0,1)}_{g \rightarrow g g},
 \nonumber \\
{\cal P}^{(1,1)}_{(1,0)\; g \rightarrow g g, sc}
 & = & 0.
\eq
Again, only the label $corr=intr$ is relevant for the applications
$e^+ e^- \rightarrow 3 \;\mbox{jets}$ and 
$e^+ e^- \rightarrow 2 \;\mbox{jets}$.
Finally for the splitting $g \rightarrow q \bar{q}$ one finds
\bq
\lefteqn{
{\cal P}^{(1,1)}_{(1,0)\; g \rightarrow q \bar{q}, lc}
 = 
  S_\eps^{-1} c_\Gamma \left( \frac{-s_{ijk}}{\mu^2} \right)^{-\eps}
   y^{-\eps}
       \left[ \frac{1}{\eps^2} + g_{1,intr}(y,z)+ g_{1,intr}(y,1-z)
 \right. } \nonumber \\
 & & \left.
              + \frac{1}{\eps^2} 
                \left(
                      \frac{1}{1-2\eps} - \frac{\eps(1-\rho\eps)}{2(1-\eps)}
                      + \frac{2\eps(1-\eps)(1-\rho\eps)}{(1-2\eps)(3-2\eps)}
                \right)
       \right] \; {\cal P}^{(0,1)}_{g \rightarrow q \bar{q}}
  - \frac{11}{6\eps} \; {\cal P}^{(0,1)}_{q \rightarrow q \bar{q}},
 \nonumber \\
\lefteqn{
{\cal P}^{(1,1)}_{(1,0)\; g \rightarrow q \bar{q}, nf}
 = 
  - S_\eps^{-1} c_\Gamma \left( \frac{-s_{ijk}}{\mu^2} \right)^{-\eps}
     y^{-\eps}
              \frac{2(1-\eps) }{\eps (1-2\eps)(3-2\eps)} 
       \; {\cal P}^{(0,1)}_{g \rightarrow q \bar{q}}
   + \frac{1}{3\eps} \; {\cal P}^{(0,1)}_{q \rightarrow q \bar{q}},
 } \nonumber \\
\lefteqn{
{\cal P}^{(1,1)}_{(1,0)\; g \rightarrow q \bar{q}, sc}
 = 
  S_\eps^{-1} c_\Gamma \left( \frac{-s_{ijk}}{\mu^2} \right)^{-\eps}
   y^{-\eps}
   \frac{1}{\eps^2}
   \left[
              \frac{1}{1-2\eps}
              - \frac{\eps(1-\rho\eps)}{2(1-\eps)} 
   \right]
       \; {\cal P}^{(0,1)}_{g \rightarrow q \bar{q}}.
}
\eq
Let us now turn to the subtraction term 
$d\alpha^{(1,1)}_{(0,1)\;n}$ and the splitting functions ${\cal P}^{(1,1)}_{(0,1)}$.
The splitting functions
${\cal P}^{(1,1)}_{(0,1)}$ are given by
\bq
\label{oneloopsplitsubtrterm}
{\cal P}^{(1,1)}_{(0,1)\; a \rightarrow b c} 
 & = &  
 \frac{S_\eps^{-1}\left(4\pi\right)^\eps}{\Gamma(1-\eps)} 
      \left( \frac{s_{ijk}}{\mu^2} \right)^{-\eps}
      \left[ h_{corr}(y,z) \right]^{-\eps} {\cal V}^{(0,1)}_{a' \rightarrow b' c'}
  \; {\cal P}^{(0,1)}_{a \rightarrow b c},
\eq
where $a' \rightarrow b' c'$ is the splitting, which has been integrated out in the subtraction
term $d\alpha^{(0,1)}_{n+1}$.
The label ``$corr$'' takes the values ``$intr\;i,j$'',``$intr\;j,k$'',``$intr\;i,k$'',
''$emit$'' or ''$spec$''. If $r=1/2$ is chosen for the subtraction terms
$d\alpha^{(0,1)}_{n+1}$, then ${\cal V}^{(0,1)}_{a' \rightarrow b' c'}$
is independent of any kinematical variable.
The function $h$ is given by
\bq
h_{intr,\;i,j}(y,z) & = & y, 
 \nonumber \\
h_{intr,\;j,k}(y,z) & = & (1-y) (1-z), 
 \nonumber \\
h_{intr,\;i,k}(y,z) & = & (1-y) z,
 \nonumber \\
h_{emit}(y,z) & = & (1-y) z \frac{2 p_e p_l}{s_{ijk}},
 \nonumber \\
h_{spec}(y,z) & = & \frac{2 p_s p_l}{s_{ijk}}.
\eq
Here, $p_l$ is the additional hard momentum appearing in eq. (\ref{approxemitspec}).


\section{Integration of the subtraction terms}
\label{sect:integrate}

The phase space measure for $n+1$ particles in $D$ dimensions factorizes according to
\bq
\int d\phi_{n+1}(P\rightarrow ..., p_i, p_j, p_k, ...)
& = & 
\int d\phi_n(P\rightarrow ..., p_e, p_s, ...)
d\phi_{unresolved}(s_{ij},s_{jk},s_{ik},\Omega_{D-2}^{(i)}).
 \nonumber \\
\eq
where $\Omega_{D-2}^{(i)}$ denotes the dependency of the unresolved phase space on the 
$D-3$ remaining angles of particle $i$ in $D$ dimensions.
In the absence of correlations between three hard partons this dependence
is trivial and
the unresolved phase space is given by
\bq
d\phi_{unresolved} 
& = & 
\frac{(4\pi)^{\eps-2}}{\Gamma\left(1-\eps\right)}  
\left( s_{ijk} \right)^{1-\eps} 
\int\limits_0^1 dy y^{-\eps} (1-y)^{1-2\eps} 
\int\limits_0^1 dz z^{-\eps} (1-z)^{-\eps}.
\eq
The variables $y$ and $z$ are given by
\bq
y = \frac{s_{ij}}{s_{ijk}}, & & 
z = \frac{s_{ik}}{s_{ik}+s_{jk}}.
\eq
The collinear limit is obtained by $y \rightarrow 0$
and in the soft limit $p_j \rightarrow 0$ one has in addition
$z \rightarrow 1$.
The splitting functions for $g \rightarrow g g$ and $g \rightarrow q \bar{q}$ involve
the spin correlation tensor $k_\perp^\mu k_\perp^\nu$.
The integral over the spin correlation tensor can be written as
\bq
\lefteqn{
\int d\phi_{unresolved} \; f(y,z) \; k_\perp^\mu k_\perp^\nu = }
 \nonumber \\
 & &
C_{21} p_e^\mu p_e^\nu + C_{22} p_s^\mu p_s^\nu
+ C_{23} \left( p_e^\mu p_s^\nu + p_s^\mu p_e^\nu \right)
+ C_{24} g^{\mu\nu}.
\eq
Using $2 k_\perp p_e = 0$, $2 k_\perp p_s = 0$, $p_e^2=0$
and $p_s^2=0$
this reduces to 
\bq
\int d\phi_{unresolved} \; f(y,z) \; k_\perp^\mu k_\perp^\nu & = &
- C_{24} \left( -g^{\mu\nu} + 2 \frac{   p_e^\mu p_s^\nu 
                                       + p_s^\mu p_e^\nu}
                                     {2 p_e p_s }
         \right).
\eq
Due to gauge invariance only the term $-C_{24} (-g^{\mu\nu})$ will give a non-vanishing
contribution.
$C_{24}$ is obtained by contracting with $g_{\mu\nu}$,
\bq
C_{24} & = & \frac{1}{2(1-\rho\eps)} \int d\phi_{unresolved} \; f(y,z) \; 
  g_{\mu\nu} k_\perp^\mu k_\perp^\nu,
\eq
and therefore
\bq
\lefteqn{
\int d\phi_{unresolved} \; f(y,z) \; k_\perp^\mu k_\perp^\nu = }
  \nonumber \\
 & &
 \frac{s_{ijk}}{2(1-\rho \eps)} \left( -g^{\mu\nu} + 2 \frac{   p_e^\mu p_s^\nu 
                                       + p_s^\mu p_e^\nu}
                                     {2 p_e p_s }
         \right)
 \int d\phi_{unresolved} \; f(y,z) \; y z (1-z).
\eq
The integration over the unresolved phase space can be reduced to three basic
integrals.
The first one,
\bq
\label{basicintegral1}
\lefteqn{
\int\limits_0^1 dy \; y^a (1-y)^{1+c+d} \int\limits_0^1 dz \; z^c (1-z)^d \left[ 1 -z(1-y)\right]^e
 = } \nonumber \\
 & & 
 \frac{\Gamma(1+a) \Gamma(1+c) \Gamma(1+d) \Gamma(2+a+d+e)}{\Gamma(2+a+d) \Gamma(3+a+c+d+e)}
\eq
is sufficient for the integration of the dipole factors ${\cal D}^{(0,1)}$.
Note that the integral in eq. (\ref{basicintegral1}) depends only on four free
parameters $a$, $c$, $d$ and $e$. The exponent of the factor $(1-y)$ is fixed
by these parameters.
At the one-loop level there are in addition integrals over the functions $g_1(y,z)$ and
$g_1(y,1-z)$, which in turn involve an integration over a hypergeometric function.
These are given by
\bq
\label{basicintegral2}
\lefteqn{
\hspace*{-1cm}
\int\limits_0^1 dy \; y^a (1-y)^{1+c+d} \int\limits_0^1 dz \; z^c (1-z)^d \left[ 1 -z(1-y)\right]^e
 {}_2F_1\left( \eps, \eps; 1+\eps; (1-y)(1-z) \right)
 = } \nonumber \\
 & & 
 \frac{\Gamma(1+a) \Gamma(1+c) \Gamma(1+\eps)}{\Gamma(\eps) \Gamma(\eps)}
 \nonumber \\
 & & \times
 \sum\limits_{j=0}^\infty 
 \frac{\Gamma(j+\eps) \Gamma(j+\eps) \Gamma(j+1+d) \Gamma(j+2+a+d+e)}
      {\Gamma(j+1) \Gamma(j+1+\eps) \Gamma(j+2+a+d) \Gamma(j+3+a+c+d+e)},
\eq
and
\bq
\label{basicintegral3}
\lefteqn{
\hspace*{-3cm}
\int\limits_0^1 dy \; y^a (1-y)^{1+c+d} \int\limits_0^1 dz \; z^c (1-z)^d \left[ 1 -z(1-y)\right]^e
 {}_2F_1\left( \eps, \eps; 1+\eps; (1-y) z \right)
 = } \nonumber \\
 & & 
 \frac{\Gamma(1+a) \Gamma(1+d) \Gamma(2+a+d+e) \Gamma(1+\eps)}{\Gamma(2+a+d) \Gamma(\eps) \Gamma(\eps)}
 \nonumber \\
 & & \times
 \sum\limits_{j=0}^\infty 
 \frac{\Gamma(j+\eps) \Gamma(j+\eps) \Gamma(j+1+c)}
      {\Gamma(j+1) \Gamma(j+1+\eps) \Gamma(j+3+a+c+d+e)},
\eq
The results are proportional to hypergeometric functions ${}_4F_3$ and ${}_3F_2$, respectively, with unit argument.
These sums are then expanded into a Laurent series in $\eps$ with the techniques
described in \cite{Moch:2001zr,Weinzierl:2002hv}.
For the symbolic calculations the 
computer algebra system GiNaC is used \cite{Bauer:2000cp}.
To simplify the resulting expressions, the following formulae are useful:
\bq
\zeta_2 = \frac{\pi^2}{6},
\;\;\;
\zeta_4 = \frac{\pi^4}{90},
\;\;\;
\zeta_{1,2} = \zeta_3,
\;\;\;
\zeta_{1,3} = \frac{\pi^4}{360},
\;\;\;
\zeta_{2,2} = \frac{\pi^4}{120},
\;\;\;
\zeta_{1,1,2} = \frac{\pi^4}{90}.
\eq
For the multiple zeta-values the notation
\bq
\zeta_{m_k,...,m_2,m_1} & = & \sum\limits_{i_1>i_2>...>i_k>0} 
               \frac{1}{i_1^{m_1} i_2^{m_2} ... i_k^{m_k} }
\eq
is used.

\subsection{Integration of the subtraction terms for single unresolved contributions}

We write the integrated dipole factors as follows:
\bq
8 \pi^2 S_\eps^{-1} \mu^{2\eps} 
\int d\phi_{unresolved} \;{\cal P}^{(0,1)}_{a \rightarrow b c}(p_i,p_j;p_k) 
 & = &
 \frac{S_\eps^{-1} (4 \pi)^{\eps}}{\Gamma(1-\eps)} \left( \frac{s_{ijk}}{\mu^2} \right)^{-\eps} 
 \; {\cal T} \; {\cal V}^{(0,1)}_{a \rightarrow b c} + \mbox{gauge terms},
 \nonumber \\
\eq
where ${\cal T}$ is the appropriate polarization sum tensor, e.g.
for quarks one has
\bq
{\cal T} &  = & \sum\limits_{\lambda} u_\lambda(p) \bar{u}_\lambda(p) = p\!\!\!/,
\eq
whereas for gluons one has
\bq
{\cal T} &  = & \sum\limits_{\lambda} \left( \eps^\mu_\lambda(k,n) \right)^\ast 
                                             \eps^\nu_\lambda(k,n)
 = - g^{\mu\nu} 
   + \frac{k^\mu n^\nu + n^\mu k^\nu}{k \cdot n} 
   - n^2 \frac{k^\mu k^\nu}{\left( k \cdot n \right)^2}.
\eq
In the case for an emitting gluon, there are additional terms indicated by ``gauge terms'', which
vanish when contracted into an amplitude.
Integration of the subtraction terms for single unresolved 
contributions yields:
\bq
\label{ressingunresolved}
{\cal V}^{(0,1)}_{q \rightarrow q g} & = & 
  \frac{\Gamma(-\eps) \Gamma(-2\eps) \Gamma(1-\eps)^2}{\Gamma(1-2\eps)\Gamma(1-3\eps)}
  \left[ 2 + \frac{4 \eps}{1-3\eps} - 2 \left(1-\rho \eps\right) \frac{\eps (1-\eps)}{(1-3\eps)(2-3\eps)} \right]
 \nonumber \\
 & = &
 \left[
          \frac{1}{\eps^2}
        + \frac{3}{2 \eps}
        + \frac{17}{4} + \frac{1}{2} \rho - \frac{1}{2} \pi^2
        + \left( \frac{99}{8} + \frac{7}{4} \rho - \frac{3}{4} \pi^2 - 8 \zeta_3 \right) \eps
 \right.
 \nonumber \\
& &
 \left.
        + \left( \frac{585}{16} + \frac{45}{8} \rho - \frac{17}{8} \pi^2 
               - \frac{1}{4} \rho \pi^2  - 12 \zeta_3
                 - \frac{11}{120} \pi^4 \right) \eps^2
 \right]
 + O(\eps^3), 
\nonumber \\
{\cal V}^{(0,1)}_{g \rightarrow g g}
 & = &
  \frac{\Gamma(-\eps) \Gamma(-2\eps) \Gamma(1-\eps)^2}{\Gamma(1-2\eps)\Gamma(1-3\eps)}
  \left[ 2 + \frac{4\eps}{1-3\eps} - \frac{4}{3}  r \frac{\eps (1-\eps)}{(1-3\eps)(2-3\eps)} \right]
 \nonumber \\
 & = &
 \left[
          \frac{1}{\eps^2}
        + \left( 2 - \frac{1}{3} r \right)\frac{1}{\eps}
        + 6 - \frac{7}{6} r - \frac{1}{2} \pi^2
        + \left( 18 - \frac{15}{4} r - \pi^2 + \frac{1}{6} r \pi^2 - 8 \zeta_3 \right) \eps
 \right.
 \nonumber \\
& &
 \left.
        + \left( 54 - \frac{93}{8} r - 3 \pi^2 
               + \frac{7}{12} r \pi^2 - 16 \zeta_3 + \frac{8}{3} r \zeta_3 
                 - \frac{11}{120} \pi^4 \right) \eps^2
 \right]
 \nonumber \\
 & &
 + O(\eps^3),
\nonumber \\
{\cal V}^{(0,1)}_{g \rightarrow q \bar{q}}
 & = &
  \frac{\Gamma(-\eps) \Gamma(1-\eps)^2}{\Gamma(2-3\eps)}
  \left[ \frac{1}{2} - \frac{2}{3} \frac{r}{(1-\rho \eps)} \frac{(1-\eps)}{(2-3\eps)} \right]
 \nonumber \\
 & = &
 \left[
        \left( -\frac{1}{2} + \frac{1}{3} r \right)\frac{1}{\eps}
        - \frac{3}{2} + \frac{7}{6} r + \frac{1}{3} \rho r
        + \left( -\frac{9}{2} + \frac{15}{4} r + \frac{3}{2} \rho r + \frac{1}{4} \pi^2 - \frac{1}{6} r \pi^2 \right) \eps
 \right.
 \nonumber \\
& &
 \left.
        + \left( -\frac{27}{2} + \frac{93}{8} r + \frac{21}{4} \rho r
               + \frac{3}{4} \pi^2 - \frac{7}{12} r \pi^2 - \frac{1}{6} \rho r \pi^2 
                 + 4 \zeta_3 - \frac{8}{3} r \zeta_3 
                 \right) \eps^2
 \right]
 \nonumber \\
 & &
 + O(\eps^3).
\eq
Since $\rho$ takes only the values $0$ and $1$, $\rho^2=\rho$ was used 
to simplify the $\eps$-expansion 
for ${\cal V}^{(0,1)}_{g \rightarrow q \bar{q}}$.

Note that these results are sandwiched in eq. (\ref{oneloopsubtrterms})
between a one-loop amplitude and a Born amplitude
and are therefore needed to ${\cal O}(\eps^3)$. 

\subsection{Integration of the subtraction terms for contributions with one virtual and one real unresolved parton}

We first consider ${\cal V}^{(1,1)}_{(1,0)}$.
We write the integrated dipole factors as follows:
\bq
 8 \pi^2 S_\eps^{-1} \mu^{2\eps}
\int d\phi_{unresolved} \;{\cal P}^{(1,1)}_{(1,0)} 
 & = &
 \frac{S_\eps^{-2} (4 \pi)^{2\eps}}{\Gamma(1-\eps)^2} \left( \frac{s_{ijk}}{\mu^2} \right)^{-2\eps} 
 \; {\cal T} \; {\cal V}^{(1,1)}_{(1,0)} + \mbox{gauge terms}.
\eq
For the splitting $q\rightarrow q g$ one finds
\bq
\lefteqn{
{\cal V}^{(1,1)}_{(1,0)\; q \rightarrow q g, lc, intr} = 
 - \frac{1}{4\eps^4}
 - \frac{31}{12 \eps^3}
 + \left( -\frac{51}{8} - \frac{1}{4} \rho + \frac{5}{12} \pi^2 - \frac{11}{6} L
   \right) \frac{1}{\eps^2} 
 + \left( - \frac{151}{6} - \frac{55}{24} \rho 
   \right.
}
 \nonumber \\
 & & 
   \left.
          + \frac{145}{72} \pi^2 + \frac{15}{2} \zeta_3
          - \frac{11}{4} L - \frac{11}{12} L^2 
   \right) \frac{1}{\eps}
 - \frac{1663}{16} - \frac{233}{24} \rho 
 + \frac{107}{16} \pi^2 + \frac{5}{12} \rho \pi^2 
 + \frac{356}{9} \zeta_3 
 \nonumber \\
 & &
 - \frac{1}{72} \pi^4
 - \frac{187}{24} L - \frac{11}{12} \rho L + \frac{55}{72} \pi^2 L
 - \frac{11}{8} L^2 - \frac{11}{36} L^3
 + i \pi \left[
            - \frac{1}{4 \eps^3}
            - \frac{3}{4 \eps^2}
            + \left( - \frac{29}{8}
              \right. \right.
 \nonumber \\
 & & \left. \left. 
                     - \frac{1}{4} \rho + \frac{\pi^2}{3} \right) \frac{1}{\eps}
            - \frac{139}{8} - \frac{11}{8} \rho + \pi^2 + \frac{15}{2} \zeta_3 
     \right]
 + {\cal O}(\eps),
 \nonumber \\
\lefteqn{
{\cal V}^{(1,1)}_{(1,0)\; q \rightarrow q g, nf} = 
 \frac{1}{3\eps^3}
 + \left( \frac{1}{2} + \frac{1}{3} L \right) \frac{1}{\eps^2}
 + \left( \frac{17}{12} + \frac{1}{6} \rho - \frac{5}{36} \pi^2 + \frac{1}{2} L 
          + \frac{1}{6} L^2 \right) \frac{1}{\eps}
}
 \nonumber \\
 & & 
 + \frac{33}{8} + \frac{7}{12} \rho - \frac{5}{24} \pi^2 - \frac{23}{9} \zeta_3
 + \frac{17}{12} L + \frac{1}{6} \rho L - \frac{5}{36} \pi^2 L
 + \frac{1}{4} L^2 + \frac{1}{18} L^3
 + {\cal O}(\eps),
 \nonumber \\
\lefteqn{
{\cal V}^{(1,1)}_{(1,0)\; q \rightarrow q g, sc} = 
 \left( \frac{5}{8} - \frac{\pi^2}{6} \right) \frac{1}{\eps^2}
 + \left( \frac{35}{8} + \frac{3}{8} \rho - \frac{\pi^2}{4} - 7 \zeta_3 \right) \frac{1}{\eps}
 + 25 + \frac{15}{4} \rho - \frac{35}{24} \pi^2 
 - \frac{1}{12} \rho \pi^2
}
 \nonumber \\
 & & 
 - \frac{21}{2} \zeta_3 - \frac{17}{90} \pi^4
 + i \pi \left[
                \left( \frac{5}{8} - \frac{\pi^2}{6} \right) \frac{1}{\eps}
                + \frac{35}{8} + \frac{3}{8} \rho
                - \frac{\pi^2}{4} - 7 \zeta_3
         \right]
 + {\cal O}(\eps).
\eq
Here, $L = \ln(s_{ijk}/\mu^2)$.
For the splitting $g\rightarrow g g$ one finds
\bq
\lefteqn{
{\cal V}^{(1,1)}_{(1,0)\; g \rightarrow g g, lc, intr} = 
 - \frac{1}{4 \eps^4}
 + \left( -\frac{17}{6} + \frac{1}{6} r \right) \frac{1}{\eps^3}
 + \left( - \frac{29}{3} + \frac{16}{9} r + \frac{7}{12} \pi^2 - \frac{11}{6} L \right) 
         \frac{1}{\eps^2}
}
 \nonumber \\
 & & 
 + \left( -43 
          + \frac{949}{108} r + \frac{199}{72} \pi^2 - \frac{1}{3} \pi^2 r
          + \frac{29}{2} \zeta_3 - \frac{11}{3} L + \frac{11}{18} r L
          - \frac{11}{12} L^2 \right) \frac{1}{\eps}
 - 193 + \frac{26875}{648} r 
 \nonumber \\
 & &
 + \frac{1}{18} \rho r 
 + \frac{415}{36} \pi^2 - \frac{499}{216} \pi^2 r 
 + \frac{1117}{18} \zeta_3 - 8 \zeta_3 r
 + \frac{7}{40} \pi^4
 - 11 L + \frac{77}{36} r L + \frac{55}{72} \pi^2 L
 - \frac{11}{6} L^2 
 \nonumber \\
 & &
 + \frac{11}{36} r L^2
 - \frac{11}{36} L^3 
 + i \pi \left[
                - \frac{1}{4 \eps^3}
                + \left( -1 + \frac{1}{6} r \right) \frac{1}{\eps^2}
                + \left( -6 + \frac{7}{6} r + \frac{\pi^2}{2} \right) \frac{1}{\eps}
                - 32 + \frac{359}{54} r 
 \right. \nonumber \\
 & & \left. 
                + \frac{5}{3} \pi^2 
                - \frac{5}{18} \pi^2 r + \frac{29}{2} \zeta_3
         \right]
 + {\cal O}(\eps),
 \nonumber \\
\lefteqn{
{\cal V}^{(1,1)}_{(1,0)\; g \rightarrow g g, nf} = 
 \frac{1}{3\eps^3}
 + \left( \frac{2}{3} - \frac{1}{9} r + \frac{1}{3} L \right) \frac{1}{\eps^2}
 + \left( 2 - \frac{2}{9} r - \frac{5}{36} \pi^2 + \frac{2}{3} L - \frac{1}{9} r L
          + \frac{1}{6} L^2 \right) \frac{1}{\eps}
}
 \nonumber \\
 & & 
 + 6 + \frac{1}{36} r + \frac{1}{9} \rho r 
 - \frac{5}{18} \pi^2 + \frac{5}{108} \pi^2 r
 - \frac{23}{9} \zeta_3 
 + 2 L - \frac{7}{18} r L - \frac{5}{36} \pi^2 L
 + \frac{1}{3} L^2 - \frac{1}{18} r L^2
 \nonumber \\
 & &
 + \frac{1}{18} L^3
 + i \pi \frac{1}{6} r
 + {\cal O}(\eps),
 \nonumber \\
\lefteqn{
{\cal V}^{(1,1)}_{(1,0)\; g \rightarrow g g, sc} = 0.
}
\eq
For the splitting $g \rightarrow q \bar{q}$ one finds
\bq
\lefteqn{
{\cal V}^{(1,1)}_{(1,0)\; g \rightarrow q \bar{q}, lc} = 
 \left( \frac{7}{8} - \frac{19}{36} r \right) \frac{1}{\eps^2}
 + \left( \frac{247}{72} + \frac{1}{24} \rho - \frac{121}{54} r - \frac{5}{9} \rho r
          - \frac{\pi^2}{6} + \frac{1}{9} \pi^2 r
          + \frac{11}{12} L 
 \right.
}
 \nonumber \\
 & & 
   \left.
          - \frac{11}{18} r L
   \right) \frac{1}{\eps}
 + \frac{3581}{216} + \frac{23}{72} \rho - \frac{7339}{648} r - \frac{329}{108} \rho r
 - \frac{119}{144} \pi^2 + \frac{139}{216} \pi^2 r + \frac{1}{9} \pi^2 \rho r
 - 7 \zeta_3 
 \nonumber \\
 & &
 + \frac{14}{3} \zeta_3 r 
 + \frac{11}{4} L - \frac{77}{36} r L - \frac{11}{18} \rho r L
 + \frac{11}{24} L^2 - \frac{11}{36} r L^2
 + i \pi \left[
                \left( - \frac{1}{24} + \frac{1}{12} r \right) \frac{1}{\eps}
                + \frac{49}{72} 
 \right. \nonumber \\
 & & \left.
                + \frac{1}{24} \rho - \frac{11}{108} r + \frac{1}{18} \rho r
                - \frac{\pi^2}{6} + \frac{1}{9} \pi^2 r 
         \right]
 + {\cal O}(\eps),
 \nonumber \\
\lefteqn{
{\cal V}^{(1,1)}_{(1,0)\; g \rightarrow q \bar{q}, nf} = 
 \left( \frac{4}{9} - \frac{7}{18} r - \frac{1}{6} L + \frac{1}{9} r L \right) 
      \frac{1}{\eps}
 + \frac{151}{54} - \frac{895}{324} r - \frac{7}{18} \rho r
 - \frac{11}{72} \pi^2 + \frac{11}{108} \pi^2 r
}
 \nonumber \\
 & & 
 - \frac{1}{2} L + \frac{7}{18} r L + \frac{1}{9} \rho r L
 - \frac{1}{12} L^2 + \frac{1}{18} r L^2
 + i \pi \left[
                \left( \frac{1}{6} - \frac{1}{9} r \right) \frac{1}{\eps}
                + \frac{17}{18} - \frac{7}{9} r - \frac{1}{9} \rho r
         \right]
 + {\cal O}(\eps),
 \nonumber \\
\lefteqn{
{\cal V}^{(1,1)}_{(1,0)\; g \rightarrow q \bar{q}, sc} = 
 \left( - \frac{1}{4} + \frac{1}{6} r \right) \frac{1}{\eps^3}
 + \left( - \frac{11}{8} + \frac{41}{36} r + \frac{1}{6} \rho r \right) \frac{1}{\eps^2}
 + \left( - \frac{51}{8} - \frac{1}{8} \rho + \frac{641}{108} r 
   \right.
}
 \nonumber \\
 & & 
   \left.
          + \frac{25}{18} \rho r
          + \frac{\pi^2}{3} - \frac{2}{9} \pi^2 r
   \right) \frac{1}{\eps}
 - \frac{219}{8} - \frac{5}{8} \rho + \frac{4441}{162} r + \frac{212}{27} \rho r
 + \frac{11}{6} \pi^2 - \frac{41}{27} \pi^2 r - \frac{2}{9} \pi^2 \rho r
 \nonumber \\
 & &
 + 5 \zeta_3 - \frac{10}{3} \zeta_3 r 
 + i \pi \left[
                 \left( - \frac{1}{4} + \frac{1}{6} r \right) \frac{1}{\eps^2}
                + \left( - \frac{11}{8} + \frac{41}{36} r + \frac{1}{6} \rho r \right) 
                  \frac{1}{\eps}
                - \frac{51}{8} - \frac{1}{8} \rho + \frac{641}{108} r
 \right. \nonumber \\
 & & \left.  
                + \frac{25}{18} \rho r
                + \frac{\pi^2}{4} - \frac{\pi^2}{6} r 
         \right]
 + {\cal O}(\eps).
\eq
Again, since $\rho$ only takes the values $0$ and $1$, $\rho^2=\rho$ was used
to simplify the results.

Let us now consider the quantities ${\cal V}^{(1,1)}_{(0,1)}$, obtained from the
integration of the subtraction term $d\alpha^{(1,1)}_{(0,1)}$.
The corresponding splitting functions ${\cal P}^{(1,1)}_{(0,1)}$ were given
in eq. (\ref{oneloopsplitsubtrterm}) by
\bq
{\cal P}^{(1,1)}_{(0,1)\; a \rightarrow b c} 
 & = &  
 \frac{S_\eps^{-1}\left(4\pi\right)^\eps}{\Gamma(1-\eps)} 
      \left( \frac{s_{ijk}}{\mu^2} \right)^{-\eps}
      \left[ h_{corr}(y,z) \right]^{-\eps} {\cal V}^{(0,1)}_{a' \rightarrow b' c'}
  \; {\cal P}^{(0,1)}_{a \rightarrow b c}.
\eq
If $r=1/2$ is chosen for the subtraction terms
$d\alpha^{(0,1)}_{n+1}$, then ${\cal V}^{(0,1)}_{a' \rightarrow b' c'}$
is independent of any kinematical variable.
In this case we define the kinematical functions ${\cal V}^{(1,1)}_{(0,1)}$ by
\bq
\label{V1101}
 8 \pi^2 S_\eps^{-1} \mu^{2\eps}
\int d\phi_{unresolved} \;{\cal P}^{(1,1)}_{(0,1)} 
 & = &
 \frac{S_\eps^{-2} (4 \pi)^{2\eps}}{\Gamma(1-\eps)^2} \left( \frac{s_{ijk}}{\mu^2} \right)^{-2\eps} 
 \; {\cal T} \; {\cal V}^{(1,1)}_{(0,1)} 
 \; {\cal V}^{(0,1)}_{a' \rightarrow b' c'}
 + \mbox{gauge terms},
 \nonumber \\
\eq
e.g. ${\cal V}^{(0,1)}_{a' \rightarrow b' c'}$ is factored out.
For two-parton correlations we have to distinguish the cases ``$intr,i,j$'',
``$intr,j,k$'' and ``$intr,i,k$''.
One finds
\bq
\lefteqn{
{\cal V}^{(1,1)}_{(0,1)\; q \rightarrow q g, intr, i,j}
 =  
 \frac{1}{3\eps^2} + \frac{3}{4\eps} 
 + \frac{11}{4} + \frac{1}{4} \rho - \frac{5}{18} \pi^2
 + \left( \frac{21}{2} + \frac{5}{4} \rho - \frac{5}{8} \pi^2 - 6 \zeta_3 \right) \eps
}
 \nonumber \\
 & &
 + \left( 41 + \frac{11}{2} \rho - \frac{55}{24} \pi^2 - \frac{5}{24} \rho \pi^2 - \frac{27}{2} \zeta_3
          - \frac{113}{1080} \pi^4 \right) \eps^2
 + {\cal O}(\eps^3),
 \nonumber \\
\lefteqn{
{\cal V}^{(1,1)}_{(0,1)\; g \rightarrow g g, intr, i,j}
 =  
 \frac{1}{3\eps^2} + \left( 1 - \frac{1}{6} r \right) \frac{1}{\eps}
 + 4 - \frac{8}{9} r - \frac{5}{18} \pi^2
 + \left( 16 - \frac{217}{54} r - \frac{5}{6} \pi^2 + \frac{5}{36} r \pi^2 
   \right.
}
 \nonumber \\
 & &
 \left. 
 - 6 \zeta_3 \right) \eps
 + \left( 64 - \frac{1379}{81} r - \frac{10}{3} \pi^2 + \frac{20}{27} r \pi^2 - 18 \zeta_3 + 3 r \zeta_3
          - \frac{113}{1080} \pi^4 \right) \eps^2
 + {\cal O}(\eps^3),
 \nonumber \\
\lefteqn{
{\cal V}^{(1,1)}_{(0,1)\; g \rightarrow q \bar{q}, intr, i,j}
 =  
 \left( - \frac{1}{4} + \frac{1}{6} r \right) \frac{1}{\eps}
 - 1 + \frac{8}{9} r + \frac{1}{6} \rho r
 + \left( -4 + \frac{217}{54} r + \frac{19}{18} \rho r + \frac{5}{24} \pi^2 
 \right.
}
 \nonumber \\
 & &
   \left.
    - \frac{5}{36} r \pi^2
   \right) \eps
 + \left( -16 + \frac{1379}{81} r + \frac{137}{27} \rho r
          + \frac{5}{6} \pi^2 - \frac{20}{27} r \pi^2 - \frac{5}{36} \rho r \pi^2
          + \frac{9}{2} \zeta_3 - 3 r \zeta_3 \right) \eps^2
 \nonumber \\
 & &
 + {\cal O}(\eps^3).
\eq
\bq
\lefteqn{
{\cal V}^{(1,1)}_{(0,1)\; q \rightarrow q g, intr, j,k}
 =  
 \frac{2}{3\eps^2}
 + \frac{3}{2\eps}
 + 6 + \frac{1}{2} \rho - \frac{5}{9} \pi^2
 + \left( 24 + 2 \rho - \frac{5}{4} \pi^2 - 12 \zeta_3 \right) \eps
}
 \nonumber \\
 & &
 + \left( 96 + 8 \rho - 5 \pi^2 - \frac{5}{12} \rho \pi^2 - 27 \zeta_3 - \frac{113}{540} \pi^4
   \right) \eps^2
 + {\cal O}(\eps^3),
 \nonumber \\
\lefteqn{
{\cal V}^{(1,1)}_{(0,1)\; g \rightarrow g g, intr, j,k}
 =  
 \frac{2}{3\eps^2}
 + \left( 2 - \frac{1}{3} r \right) \frac{1}{\eps}
 + 8 - \frac{13}{9} r - \frac{5}{9} \pi^2 
 + \left( 32 - \frac{160}{27} r - \frac{5}{3} \pi^2 + \frac{5}{18} r \pi^2 
 \right.
}
 \nonumber \\
 & &
 \left. 
 - 12 \zeta_3 \right) \eps
 + \left( 128 - \frac{1936}{81} r - \frac{20}{3} \pi^2
          + \frac{65}{54} r \pi^2 - 36 \zeta_3 + 6 r \zeta_3 - \frac{113}{540} \pi^4 \right) \eps^2
 + {\cal O}(\eps^3),
 \nonumber \\
\lefteqn{
{\cal V}^{(1,1)}_{(0,1)\; g \rightarrow q \bar{q}, intr, j,k}
 =  
 \left( - \frac{1}{2} + \frac{1}{3} r \right) \frac{1}{\eps}
 - 2 + \frac{13}{9} r + \frac{1}{3} \rho r
 + \left( - 8 + \frac{160}{27} r + \frac{16}{9} \rho r + \frac{5}{12} \pi^2 
 \right.
}
 \nonumber \\
 & &
 \left.
 - \frac{5}{18} r \pi^2
   \right) \eps
 + \left( -32  + \frac{1936}{81} r + \frac{208}{27} \rho r
          + \frac{5}{3} \pi^2 - \frac{65}{54} r \pi^2 - \frac{5}{18} \rho r \pi^2
          + 9 \zeta_3 - 6 r \zeta_3 \right) \eps^2
 \nonumber \\
 & &
 + {\cal O}(\eps^3).
\eq
\bq
\lefteqn{
{\cal V}^{(1,1)}_{(0,1)\; q \rightarrow q g, intr, i,k}
 =  
 \frac{1}{\eps^2}
 + \frac{3}{2\eps}
 + \frac{11}{2} + \frac{1}{2} \rho - \frac{5}{6} \pi^2
 + \left( 21 + \frac{5}{2} \rho - \frac{5}{4} \pi^2 - 18 \zeta_3 \right) \eps
}
 \nonumber \\
 & &
 + \left( 82 + 11 \rho - \frac{55}{12} \pi^2 - \frac{5}{12} \rho \pi^2
          - 27 \zeta_3 - \frac{113}{360} \pi^4 \right) \eps^2
 + {\cal O}(\eps^3),
 \nonumber \\
\lefteqn{
{\cal V}^{(1,1)}_{(0,1)\; g \rightarrow g g, intr, i,k}
 =  
 \frac{1}{\eps^2}
 + \left( 2 -\frac{1}{3} r \right) \frac{1}{\eps}
 + 8 - \frac{13}{9} r - \frac{5}{6} \pi^2
 + \left( 32 - \frac{160}{27} r - \frac{5}{3} \pi^2 + \frac{5}{18} r \pi^2 
 \right.
}
 \nonumber \\
 & &
 \left.
 - 18 \zeta_3 \right) \eps
 + \left( 128 - \frac{1936}{81} r - \frac{20}{3} \pi^2 + \frac{65}{54} r \pi^2
          - 36 \zeta_3 + 6 r \zeta_3 - \frac{113}{360} \pi^4 \right) \eps^2
 + {\cal O}(\eps^3),
 \nonumber \\
\lefteqn{
{\cal V}^{(1,1)}_{(0,1)\; g \rightarrow q \bar{q}, intr, i,k}
 =  
 \left( - \frac{1}{2} + \frac{1}{3} r \right) \frac{1}{\eps}
 - 2 + \frac{13}{9} r + \frac{1}{3} \rho r
 + \left( - 8 + \frac{160}{27} r + \frac{16}{9} \rho r
        + \frac{5}{12} \pi^2 
 \right.
}
 \nonumber \\
 & &
 \left.
 - \frac{5}{18} r \pi^2 \right) \eps
 + \left( -32 + \frac{1936}{81} r + \frac{208}{27} \rho r 
          + \frac{5}{3} \pi^2 - \frac{65}{54} r \pi^2 - \frac{5}{18} \rho r \pi^2
          + 9 \zeta_3 - 6 r \zeta_3 \right) \eps^2
 \nonumber \\
 & &
 + {\cal O}(\eps^3).
\eq
For the NNLO calculation for 
$e^+ e^- \rightarrow 3 \;\mbox{jets}$
it remains to treat the cases ${\cal V}^{(1,1)}_{(0,1)\; emit}$
and ${\cal V}^{(1,1)}_{(0,1)\; spec}$.
These are relatively simple and given by

\bq
{\cal V}^{(1,1)}_{(0,1)\; a \rightarrow b c, emit} & = & 
  \left( \frac{2 p_e p_l}{s_{ijk}}\right)^{-\eps} 
  {\cal V}^{(1,1)}_{(0,1)\; a \rightarrow b c, intr, i, k},
 \nonumber \\
{\cal V}^{(1,1)}_{(0,1)\; a \rightarrow b c, spec} & = & 
  \left( \frac{2 p_s p_l}{s_{ijk}} \right)^{-\eps} 
  {\cal V}^{(0,1)}_{a \rightarrow b c}.
\eq
If the variable $r$ in ${\cal V}^{(0,1)}_{a' \rightarrow b' c'}$
is chosen according to eq. (\ref{choice2r}), then in general
${\cal V}^{(0,1)}_{a' \rightarrow b' c'}$ will depend on kinematical
invariants.
This implies that ${\cal V}^{(0,1)}_{a' \rightarrow b' c'}$
cannot be factored out from the integration over the unresolved 
phase space, as it was done in eq. (\ref{V1101}).
The results for this case are not listed here, but can be obtained easily.


\section{Summary and conclusions}
\label{sect:concl}

In this paper I considered the subtraction terms for one-loop amplitudes
in the case where two partons become collinear or where 
one parton becomes soft.
These subtraction terms are relevant for precision calculations at 
next-to-next-to-leading order.
For two- and three-jet production in electron-positron annihilation 
all necessary subtraction terms have been given and 
the subtraction terms have been integrated analytically over the
unresolved phase space.
Once the corresponding subtraction terms for double unresolved 
configurations have been worked out, the method for the cancellation
of infrared singularities is complete and it should be
possible to extend the exisiting numerical programs for NLO predictions
on $e^+ e^- \rightarrow 4 \;\mbox{jets}$ 
\cite{Dixon:1997th}-\cite{Weinzierl:1999yf}
towards NNLO predictions
for $e^+ e^- \rightarrow 3 \;\mbox{jets}$.



\end{document}